\documentclass[onecolumn]{elsart3p}
\usepackage{epsfig}
\usepackage{amssymb}
\usepackage{psfrag}
\usepackage{rotating}

\begin{document}
\newcommand{\be}{\begin{equation}}
\newcommand{\ee}{\end{equation}}
\newcommand{\ba}{\begin{eqnarray}}
\newcommand{\ea}{\end{eqnarray}}
\newcommand{\ol}{\overline}
\newcommand{\grad}{\nabla}
\newcommand{\la}{\langle}
\newcommand{\ra}{\rangle}
\newcommand{\Par}{\parallel}

\begin{frontmatter}

\title{Preserving Monotonicity in Anisotropic Diffusion}
\author[lab1,lab2]{Prateek Sharma\corauthref{cor1}}
\ead{psharma@astro.berkeley.edu}
\author[lab2]{\& Gregory W. Hammett}
\corauth[cor1]{Corresponding author. Tel.: +1 510 642 2359;
fax: +1 510 642 3411.}
\address[lab1]{Astronomy Department, University of California, Berkeley, CA 94720}
\address[lab2]{Princeton Plasma Physics Laboratory, Forrestal Campus, Princeton,
NJ 08543}

\begin{abstract}
We show that standard algorithms for anisotropic diffusion based on
centered differencing (including the recent symmetric algorithm) do
not preserve monotonicity. In the context of anisotropic thermal
conduction, this can lead to the violation of the entropy
constraints of the second law of thermodynamics, causing heat to
flow from regions of lower temperature to higher temperature. In
regions of large temperature variations, this can cause the
temperature to become negative.  Test cases to illustrate this for
centered asymmetric and symmetric differencing are presented.
Algorithms based on slope limiters, analogous to those used in
second order schemes for hyperbolic equations, are proposed to fix
these problems. While centered algorithms may be good for many
cases, the main advantage of limited methods is that they are
guaranteed to avoid negative temperature (which can cause numerical
instabilities) in the presence of large temperature gradients.  In
particular, limited methods will be useful to simulate hot, dilute
astrophysical plasmas where conduction is anisotropic and the
temperature gradients are enormous, e.g., collisionless shocks and
disk-corona interface.
\end{abstract}


\end{frontmatter}

\section{Introduction}
\label{sec:introduction}
Anisotropic diffusion, in which the rate of diffusion of some quantity is
faster in certain directions than others, occurs in many different physical
systems and applications.  Examples include diffusion in geological
formations~\cite{Saadatfar2002}, thermal properties of structural materials
and crystals~\cite{Dian-lin1991}, image
processing~\cite{Perona1990,Caselles1998,Mrazek2001},
biological systems, and plasma physics. Diffusion Tensor Magnetic Resonance
Imaging makes use of
anisotropic diffusion to distinguish different types of tissue as a medical
diagnostic~\cite{Basser2002}.  In plasma physics, the collision operator gives
rise to anisotropic diffusion in velocity space, as does the quasilinear
operator describing the interaction of particles with waves~\cite{Stix1992}.
In magnetized plasmas, thermal conduction can be much more rapid along the
magnetic field line than across it; this will be the main application in
mind for this paper.

Centered finite differencing is commonly used to implement anisotropic
thermal conduction in fusion and astrophysical
plasmas~\cite{Gunter2005,Parrish2005,Sharma2006}. Methods based on finite
differencing~\cite{Gunter2005} and higher order finite
elements~\cite{Sovinec2004} are able to simulate highly anisotropic
thermal conduction~($\chi_\parallel/\chi_\perp \sim 10^9$,
where $\chi_\parallel$ and $\chi_\perp$ are parallel and
perpendicular conduction coefficients, respectively) in laboratory plasmas.
``Symmetric" differencing introduced in~\cite{Gunter2005} is
particularly simple and has some desirable properties: perpendicular
numerical diffusion is independent of parallel conduction coefficient
$\chi_\parallel$, perpendicular numerical diffusion is small, and the
numerical heat flux operator is self adjoint. While in the symmetric method
the components of the heat flux are located at cell corners, they are
located at the cell faces in the ``asymmetric" method. The asymmetric method
has been used to study convection in anisotropically conducting
plasmas~\cite{Parrish2005} and in simulations of collisionless accretion
disks~\cite{Sharma2006}.

An important fact that has been overlooked is that the methods based
on centered differencing can give heat fluxes inconsistent with the
second law of thermodynamics, i.e., heat can flow from lower to
higher temperatures. This accentuates temperature extrema and may
result in negative temperatures at some grid points, causing
numerical instabilities as the sound speed becomes imaginary. Also,
in image processing applications it is required that no new spurious
extrema are generated with anisotropic diffusion~\cite{Perona1990},
making centered differencing unviable.

We show that both the symmetric and asymmetric methods can be modified so
that temperature extrema are not accentuated.
The components of the anisotropic heat flux consist of two contributions:
the normal term
and the transverse term (see \S 2).
The normal term for the asymmetric method (like isotropic conduction)
always gives heat flux from higher to lower temperatures,
but the transverse
term can be of any sign. The transverse term can be ``limited"
to ensure that temperature extrema are not accentuated. We use slope
limiters, analogous to those used in second order methods for
hyperbolic problems \cite{VanLeer1979,Leveque2002}, to limit the
transverse heat fluxes. For the symmetric method, where
primary heat fluxes are located at cell corners,  both normal and transverse
terms need to be limited.
Limiting based on the entropy-like function ($\dot{s}^* \equiv -\vec{q}
\cdot \vec{\nabla} T \geq 0$) is also discussed.

Limiting introduces numerical diffusion in the perpendicular
direction, and the desirable property of the symmetric method that
perpendicular pollution is independent of $\chi_\parallel$ no longer
holds. The ratio of perpendicular numerical diffusion and the
physical parallel conductivity with a Monotonized Central (MC; see
\cite{Leveque2002} for a discussion of slope limiters) limiter is
$\chi_{\perp, {\rm num}} / \chi_\parallel \sim 10^{-3}$ for a modest
number of grid points ($\sim 100$ in each direction). This clearly
is not adequate for simulating laboratory plasmas which require
$\chi_\parallel/\chi_\perp \sim 10^9$ because perpendicular numerical
diffusion will swamp the true perpendicular diffusion. For
laboratory plasmas the temperature profile is relatively smooth and
the negative temperature problem does not arise, so symmetric
differencing \cite{Gunter2005} or higher order finite elements
\cite{Sovinec2004} may be adequate. However, astrophysical plasmas
can have sharp temperature gradients, e.g., the transition region of
the sun separating the hot corona and the much cooler chromosphere, or
the disk-corona interface in accretion flows. In these applications
centered differencing may lead to negative temperatures giving
rise to numerical instabilities.
Limiting introduces somewhat larger  perpendicular numerical diffusion
but will ensure that heat flows in the correct direction at temperature
extrema; hence negative temperatures are avoided. Even a modest anisotropy
in conduction~($\chi_\parallel/\chi_\perp \lesssim 10^3$) should be enough
to study the qualitatively new effects of anisotropic conduction on
dilute astrophysical plasmas~\cite{Parrish2005}, but the positivity of
temperature is absolutely essential for numerical robustness.

The paper is organized as follows. In \S 2 we describe the heat
equation with anisotropic conduction and its numerical
implementation using asymmetric and symmetric centered differencing.
In \S 3 we present simple test problems for which centered
differencing results in negative temperatures. Limiting as a method
to avoid unphysical behavior at temperature extrema is introduced in
\S 4 \& \S 5. Slope limiters are discussed in \S 4 and limiting
based on the entropy-like condition in \S 5. Some mathematical
properties of limited methods are discussed in \S 6. In \S 7 we
compare different methods and their convergence properties with some
test problems. We conclude in \S 8.

\section{Anisotropic thermal conduction}
Thermal conduction in plasmas with the mean free path much larger than
the gyroradius is anisotropic with respect to the magnetic field
lines; heat flows primarily along the field lines
with little conduction in the perpendicular direction~\cite{Braginskii1965}.
In such cases, a divergence of anisotropic heat flux is added to
the energy equation. Thermal conduction can modify the characteristic
structure of the magnetohydrodynamic (MHD) equations making it difficult
to incorporate into upwind methods. However, thermal conduction can be
evolved  independently of the MHD equations using operator splitting,
as done in~\cite{Parrish2005}.
The equation for the evolution of internal energy density due to
anisotropic thermal conduction is \ba \label{eq:anisotropic_conduction}
\frac{\partial e}{\partial t} &=& - \vec{\nabla} \cdot \vec{q}, \\
 \vec{q} &=& - \vec{b} n (\chi_\Par-\chi_\perp)
\nabla_\parallel T - n \chi_\perp \vec{\nabla} T, \ea where $e$ is
the internal energy per unit volume, $\vec{q}$ is the heat flux,
$\chi_\Par$ and $\chi_\perp$ are the coefficients of parallel and
perpendicular conduction with respect to the local field
direction~(with dimensions $L^2T^{-1}$), $n$ is the number density,
$T \equiv (\gamma-1)e/n$ is the temperature with $\gamma=5/3$ as the
ratio of specific heats for an ideal gas, $\vec{b}$ is the unit
vector along the field line, and $\nabla_\parallel=\vec{b}
\cdot \vec{\nabla}$ represents the derivative along the magnetic field
direction. Throughout the paper we use $\gamma=2$ to avoid factors of
$2/3$ and $5/3$; results of the paper are not affected by this choice.
\begin{figure}
\centering
\includegraphics[width=3in,height=3in]{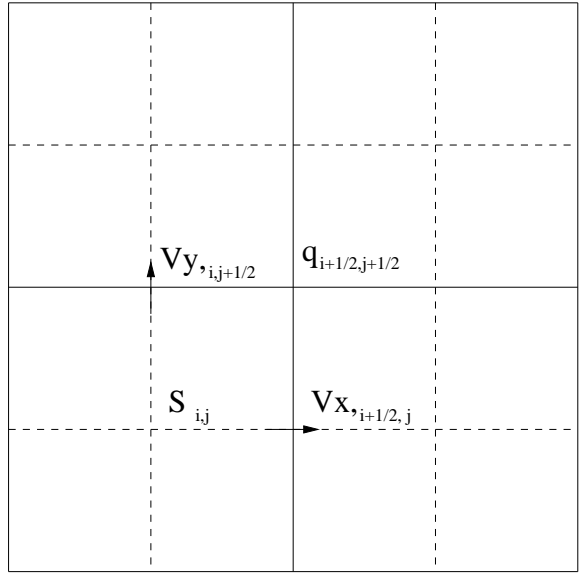}
\caption{A staggered grid with scalars $S_{i,j}$ (e.g., $n$, $e$,
and $T$) located at cell centers. The components of vectors, e.g.,
$\vec{b}$ and $\vec{q}$, are located at cell faces. Note, however,
that for the symmetric method the primary heat fluxes are located at
the cell corners~\cite{Gunter2005}, and the face centered flux is
obtained by interpolation (see \S 2.2).\label{fig:fig1}}
\end{figure}

We consider a staggered grid with the scalars like $n$, $e$, and $T$
located at the cell centers and the components of vectors, e.g.,
$\vec{b}$ and $\vec{q}$, located at the cell faces~\cite{Stone1992},
as shown in Figure \ref{fig:fig1}. The face centered components
of vectors naturally represent the flux of scalars out of a cell.
All the methods presented here are conservative and fully explicit.
It should be possible to take longer time steps with an implicit
generalization of the schemes discussed in the paper, but the
construction of fast implicit schemes for anisotropic conduction is
non-trivial.

In two dimensions the internal energy density is updated as follows,
\be
\label{eq:e_evolve} e^{n+1}_{i,j} = e^{n}_{i,j} - \Delta t \left[
\frac{q^n_{x,i+1/2,j}-q^n_{x,i-1/2,j}}{\Delta x} +
\frac{q^n_{y,i,j+1/2}-q^n_{y,i,j-1/2}}{\Delta y} \right], \ee where
the time step $\Delta t$, satisfies the stability condition 
\cite{Richtmyer1967} (ignoring density variations)
\be
\label{eq:TimeStep} \Delta t \leq \frac{\mbox{min}[\Delta x^2,
\Delta y^2]}{2(\chi_\parallel + \chi_\perp)}, \ee $\Delta x $
and $\Delta y$ are grid sizes in the two directions. The generalization
to three dimensions is straightforward.

The methods we discuss differ in the way heat fluxes are calculated
at the faces. In rest of the section we discuss the methods based on
asymmetric and symmetric centered differencing as discussed in
\cite{Gunter2005}. From here on  $\chi$ will represent parallel
conduction coefficient in cases where an explicit perpendicular
diffusion is not considered~(i.e., the only perpendicular diffusion
is due to numerical effects).

\subsection{Centered asymmetric scheme}
\begin{figure}
\centering
\psfrag{A}{\large{$(n\chi)_{-1/2}$}}
\psfrag{B}{\large{$(n\chi)_{1/2}$}} \psfrag{C}{\large{$T_0$}}
\psfrag{D}{\large{$T_{1}$}} \psfrag{E}{\large{$T_{-1}$}}
\includegraphics[width=3in,height=3in]{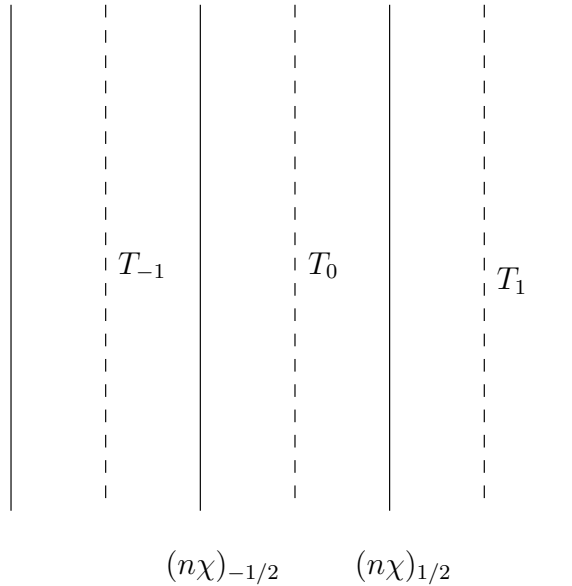}
\caption{This figure provides a motivation for using a harmonic
average for $\ol{n}\ol{\chi}$. Consider a 1-D case with the
temperatures and $n\chi$'s as shown in the figure. Given $T_{-1}$
and $T_1$, and the $n\chi$'s at the faces, we want to calculate an
average $\ol{n}\ol{\chi}$ between cells $-1$ and $1$. Assumption of
a constant heat flux gives, $q_{-1/2}=q_{1/2}=\ol{q}$, i.e.,
$-(n\chi)_{-1/2} (T_0-T_{-1})/\Delta x = -(n\chi)_{1/2}
(T_1-T_0)/\Delta x = -\ol{n}\ol{\chi} (T_1-T_{-1})/ 2 \Delta x$.
This immediately gives a harmonic mean, which is weighted towards
the smaller of the two arguments, for the interpolation
$\ol{n}\ol{\chi}$. \label{fig:fig2}}
\end{figure}

The heat flux in the $x$- direction (in 2-D), using the asymmetric
method is given by \be \label{eq:q1_asymmetric} q_{x,i+1/2,j}=-
\overline{n}\overline{\chi}  b_x \left[ b_x \frac{\partial
T}{\partial x} + \overline{b_y} \overline{\frac{\partial T}{\partial
y}} \right], \ee where overline represents the variables
interpolated to the face at $(i+1/2,j)$. The variables without an
overline are naturally located at the face.
The interpolated quantities at the face are given by simple arithmetic
averaging, \ba
\label{eq:asymmetric_bavg} \overline{b_y} &=&
(b_{y,i,j-1/2}+b_{y,i+1,j-1/2}
+ b_{y,i,j+1/2}+b_{y,i+1,j+1/2})/4, \\
\label{eq:asymmetric_Tavg} \overline{\partial T/\partial y} &=&
(T_{i,j+1}+T_{i+1,j+1}-T_{i,j-1}-T_{i+1,j-1})/4 \Delta y. \ea

We use a harmonic mean to interpolate the product of number density
and conductivity \cite{Hyman2002}, \be \label{eq:asymmetric_navg}
\frac{2}{\ol{n}\ol{\chi}}=\frac{1}{(n\chi)_{i,j}}+\frac{1}{(n\chi)_{i+1,j}};
\ee this is second order accurate for smooth regions, but
$\ol{n}\ol{\chi}$ becomes proportional to the minimum of the two
$n\chi$'s on either side of the face when the two differ
significantly. Figure \ref{fig:fig2} gives the motivation for using
a harmonic average. 
Physically, using a harmonic average preserves the robust result that the
heat flux into a region should go to zero as the density in that region
goes to zero, as in a thermos bottle using a vacuum for insulation.
Harmonic averaging is also necessary for the
method to be stable with the time step in Eq. (\ref{eq:TimeStep}).
Instead, if we use a simple mean, the stable time step condition
becomes severe by a factor $ \sim
\mbox{max}[n_{i+1,j},n_{i,j}]/2\mbox{min} [n_{i+1,j}, n_{i,j}]$,
which can result in an unacceptably small time step for initial
conditions with a large density contrast. Physically, this is
because the heat capacity is very small in low density regions, so
even a tiny heat flux into that region causes rapid changes in
temperature. Analogous expressions can be written for heat flux in
other directions.

\subsection{Centered symmetric scheme}
The notion of symmetric differencing was introduced in
\cite{Gunter2005}, where primary heat fluxes are located at the cell
corners, with \be \label{eq:q1_symmetric} q_{x,i+1/2,j+1/2} =
-\overline{n}\overline{\chi} \overline{b_x} \left [ \overline{b_x}
\overline{\frac{\partial T}{\partial x}} + \overline{b_y}
\overline{\frac{\partial T}{\partial y}}   \right ], \ee where
overline represents the interpolation of variables at the corner
given by a simple  arithmetic average,
\ba
\label{eq:symmetric_bxavg}
\overline{b_x} &=& (b_{x,i+1/2,j}+b_{x,i+1/2,j+1})/2, \\
\label{eq:symmetric_byavg}
\overline{b_y} &=& (b_{y,i,j+1/2}+b_{y,i+1,j+1/2})/2, \\
\label{eq:symmetric_Tavg} \overline{\partial T/\partial x} &=&
(T_{i+1,j}+T_{i+1,j+1}-T_{i,j}-T_{i,j+1})/2 \Delta x, \\
\overline{\partial T/\partial y} &=&
(T_{i,j+1}+T_{i+1,j+1}-T_{i,j}-T_{i+1,j})/2 \Delta y. \ea As before
(and for the same reasons), a harmonic average is used for the
interpolation of $n\chi$,
\be \label{eq:symmetric_navg}
\frac{4}{\ol{n}\ol{\chi}}= \frac{1}{(n\chi)_{i,j}} +
\frac{1}{(n\chi)_{i+1,j}} +
\frac{1}{(n\chi)_{i,j+1}}+\frac{1}{(n\chi)_{i+1,j+1}}. \ee 
Analogous expressions can be written for $q_{y,i+1/2,j+1/2}$.  The harmonic
average here is different from~\cite{Gunter2005}, who use an arithmetic average.  
Ref.~\cite{Gunter2005}
is primarily interested in magnetic fusion applications, where density
variations are usually well resolved (shocks are usually not important in
magnetic fusion) so arithmetic averaging will work well.  But there might
be some magnetic fusion cases, such as instabilities in the edge region of
a fusion device, where there might be large density variations per grid
cell and a harmonic average could be useful.  All of the test cases in~\cite{Gunter2005}
used a uniform density and so will not be affected by the choice of
arithmetic or harmonic average.

The heat fluxes located at the cell faces, $q_{x,i+1/2,j}$ and
$q_{y,i,j+1/2}$, to be used in Eq.~(\ref{eq:e_evolve}) are given
by an arithmetic average, \ba
q_{x,i+1/2,j} &=& (q_{x,i+1/2,j+1/2}+q_{x,i+1/2,j-1/2})/2, \\
q_{y,i,j+1/2} &=& (q_{y,i+1/2,j+1/2}+q_{y,i-1/2,j+1/2})/2.\ea As
demonstrated in \cite{Gunter2005}, the symmetric heat flux satisfies
the self adjointness property (equivalent to $\dot{s}^* \equiv -  \vec{q}
\cdot \vec{\nabla} T \geq 0$) at cell corners and has the desirable
property that the perpendicular numerical diffusion
($\chi_{\perp,{\rm num}}$) is independent of $\chi_\parallel/\chi_\perp$ (see
Figure 6 in~\cite{Gunter2005}). But, as we show later, both symmetric
and asymmetric schemes do not satisfy the crucial local property
that heat must flow from higher to lower temperatures, the violation
of which may result in negative temperature with large temperature
gradients.

The heat flux in the $x$- direction $q_x$
consists of two terms: the normal term $q_{xx}=-n \chi b_x^2
\partial T/\partial x$ and the transverse term $q_{xy}=-n \chi b_x
b_y \partial T/\partial y$. The asymmetric scheme uses a 2 point
stencil to calculate the normal gradient and a 6 point stencil to
calculate the transverse gradient, as compared to the symmetric
method that uses a 6 point stencil for both (hence the name
symmetric). This makes the symmetric method less sensitive to the
orientation of coordinate system with respect to the field lines.
\begin{figure}
\centering
\includegraphics[width=2.95in,height=2.5in]{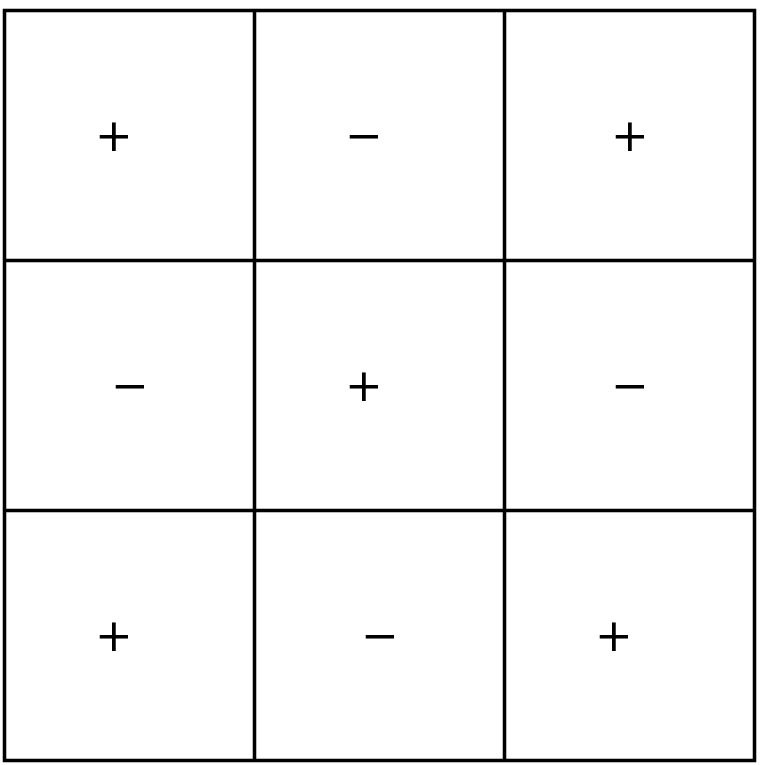}
\caption{The symmetric method is unable to diffuse a temperature
distributed in a chess-board pattern. The plus ($+$) and minus ($-$)
symbols denote two unequal temperatures.  The average of $\partial
T/\partial x|_{i+1/2,j}=(T_+ - T_-)/\Delta x$ and $\partial
T/\partial x|_{i+1/2,j+1}=(T_- - T_+)/\Delta x$ to calculate
$\partial T/\partial x|_{i+1/2,j+1/2}= \partial T/\partial
x|_{i+1/2,j} + \partial T/\partial x|_{i+1/2,j+1}$ vanishes, similarly
$\partial T/\partial y|_{i+1/2,j+1/2}=0$. \label{fig:fig3}}
\end{figure}

A problem with the symmetric method which is immediately apparent is
its inability to diffuse away a chess-board temperature pattern as
$\overline{\partial T/\partial x}$ and $\overline{\partial
T/\partial y}$, located at the cell corners, vanish for this initial
condition (see Figure \ref{fig:fig3}).

\section{Negative temperature with centered differencing}
\label{sec:Negative} In this section we present two simple test
problems that demonstrate that negative temperatures can arise
with both asymmetric and symmetric centered differencing.

\subsection{Asymmetric method}

\begin{figure}
\centering
\includegraphics[width=3in,height=3in]{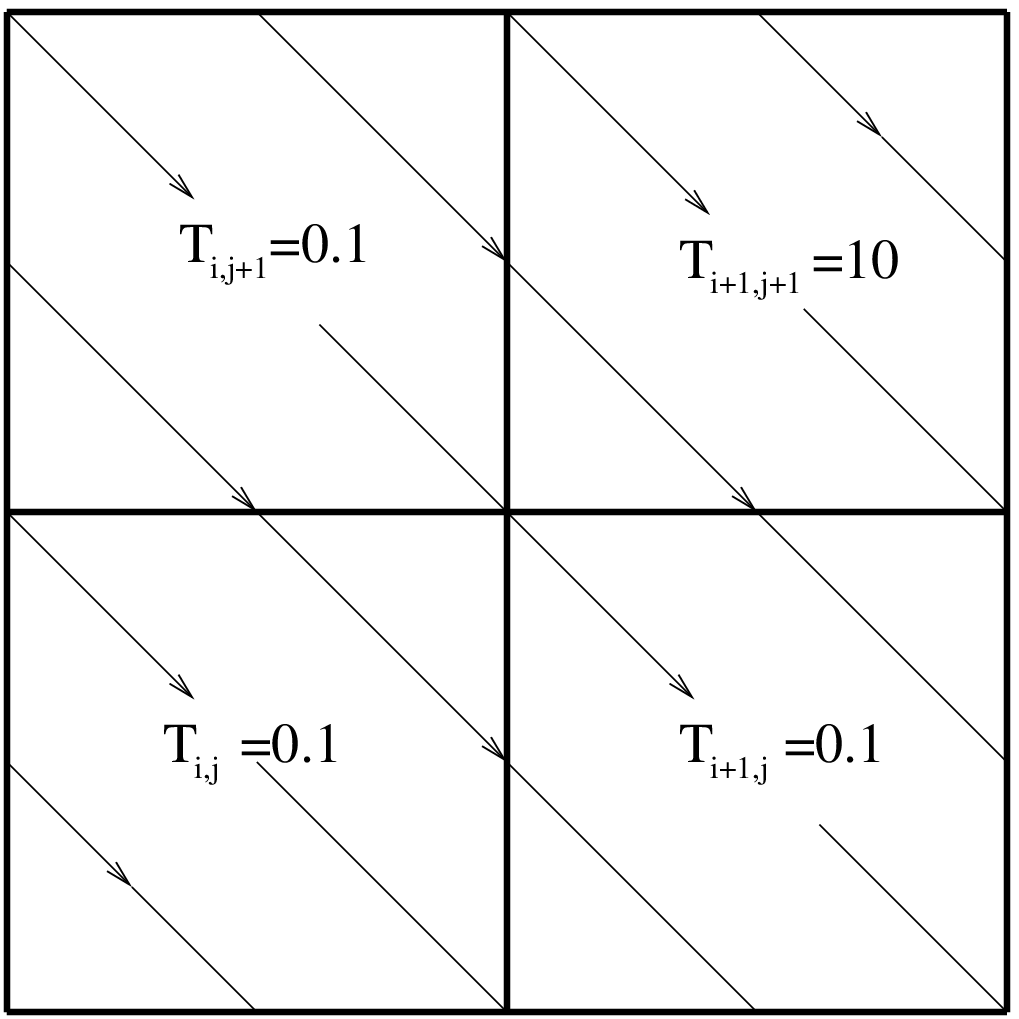}
\includegraphics[width=5in,height=4in]{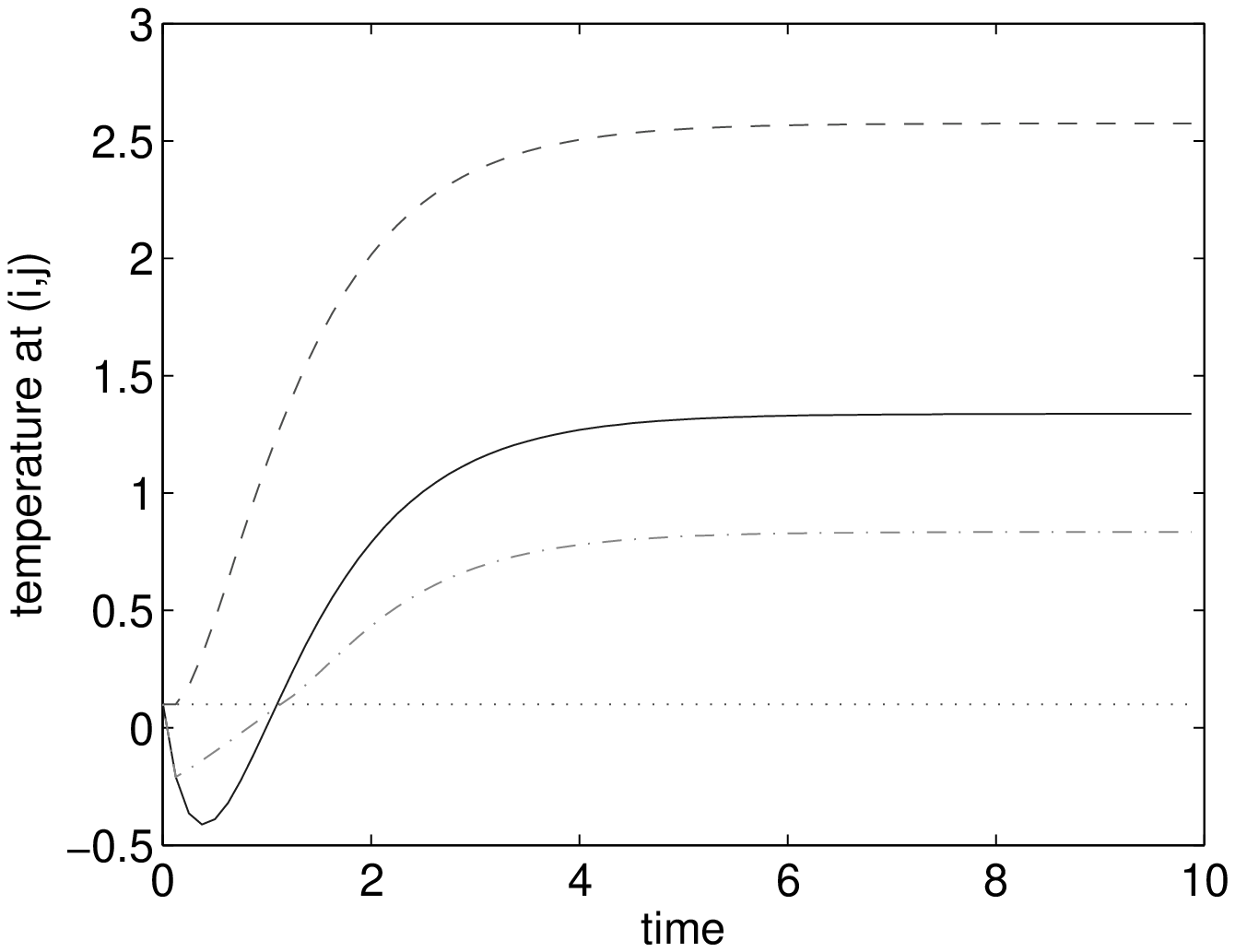}
\caption{Test problem to show that the asymmetric method can
result in negative temperature. Magnetic field lines are along
the diagonal with $b_x=-b_y=1/\sqrt{2}$. With the asymmetric method
heat flows out of the third quadrant which is already a temperature
minimum, resulting in a negative temperature $T_{i,j}$.
However due to numerical perpendicular diffusion, at late times the
temperature becomes positive again. The temperature at $(i,j)$ is
shown for different methods: asymmetric (solid line), symmetric
(dotted line), asymmetric and symmetric with slope limiters (dashed
line; both give the same result), and symmetric with entropy
limiting (dot dashed line).\label{fig:fig4}}
\end{figure}

Consider a $2 \times 2$  grid with a hot zone ($T=10$) in the first
quadrant and cold temperature ($T=0.1$) in the rest, as shown in
Figure \ref{fig:fig4}. Magnetic field is uniform over the box
with $b_x=-b_y=1/\sqrt{2}$. Number density is a constant equal to
unity. Reflecting boundary conditions are used for temperature. Using 
the asymmetric
scheme for heat fluxes out of the grid point $(i,j)$ (the third
quadrant) gives, $q_{x,i-1/2,j}=q_{y,i,j-1/2}=0$, and $q_{x,i+1/2,j}
= q_{y,i,j+1/2} = (9.9/8) n \chi/\Delta x$ (where $\Delta x=\Delta
y$ is assumed). Thus, heat flows out of the grid point $(i,j)$, which is
already a temperature minimum. This results in the temperature
becoming negative. Figure \ref{fig:fig4} shows the temperature in
the third quadrant vs.\ time for different methods. The asymmetric
method gives negative temperature ($T_{i,j}<0$) for first few time
steps, which eventually becomes positive. All other methods (except
the one based on entropy limiting) give positive temperatures at all
times for this problem. Methods based on limited
temperature gradients will be discussed later. This test
demonstrates that the asymmetric method may not be suitable for problems
with large temperature gradients because negative temperature results in
numerical instabilities.

\subsection{Symmetric method}
\label{subsec:symmfail}
\begin{figure}
\centering
\includegraphics[width=3in,height=3in]{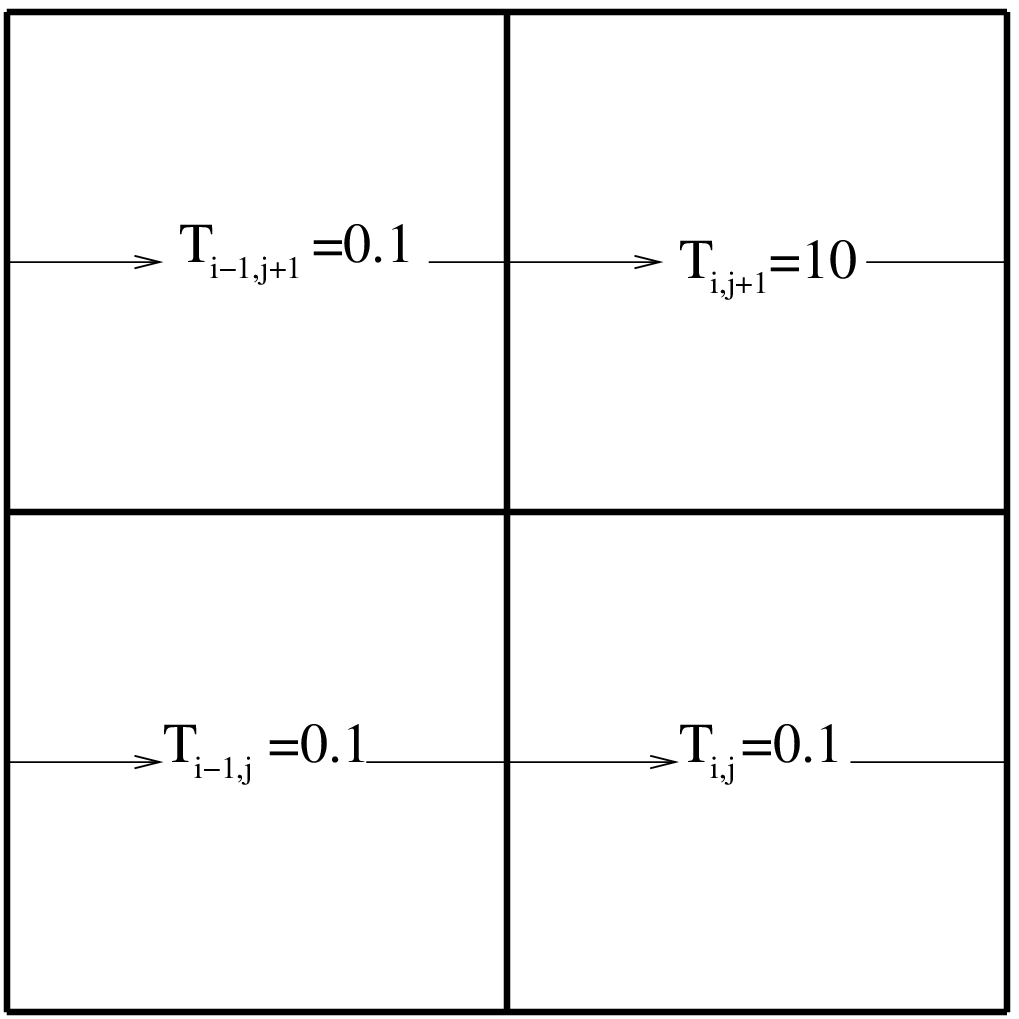}
\includegraphics[width=5in,height=4in]{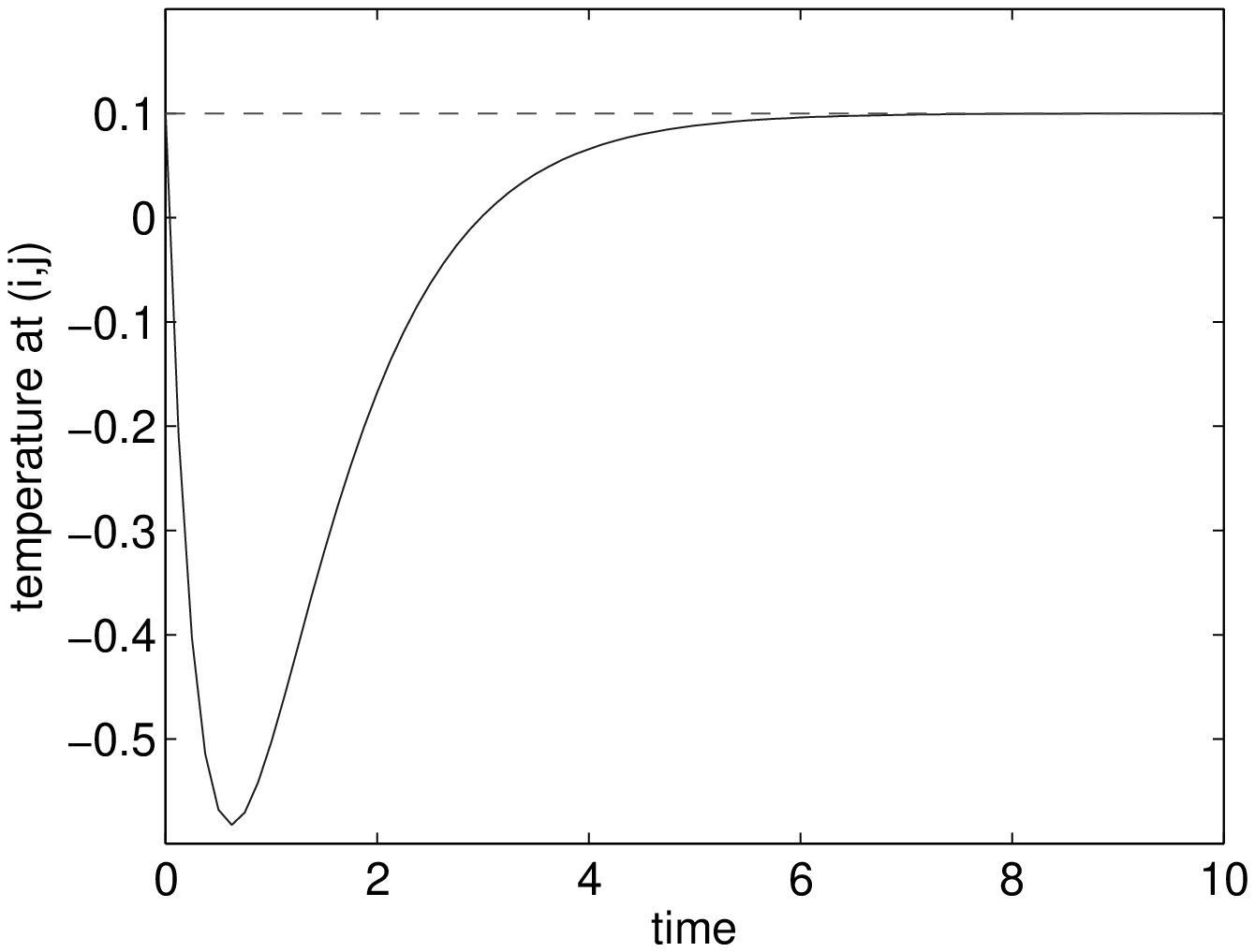}
\caption{Test problem for which the symmetric method gives negative
temperature at $(i,j)$. Magnetic field is
along the $x$- direction, $b_x=1$ and $b_y=0$. With this initial
condition, all heat fluxes into $(i,j)$ should vanish and the
temperature $T_{i,j}$ should not evolve. All methods except the
symmetric method (asymmetric, and slope and entropy limited methods)
give a constant temperature $T_{i,j}=0.1$ at all times. But with the
symmetric method, the temperature at $(i,j)$ becomes negative due to
the heat flux out of the corner $(i-1/2,j+1/2)$. The temperature
$T_{i,j}$ eventually becomes equal to the initial value of $0.1$.
\label{fig:fig5}}
\end{figure}
The symmetric method does not give negative temperature with the
test problem of the previous section. In fact, the symmetric method
gives the correct result for temperature with no numerical diffusion
in the perpendicular direction (zero heat flux out of the grid point
$(i,j)$, see Figure \ref{fig:fig4}). Other methods resulted in a
temperature increase at $(i,j)$ because of perpendicular numerical
diffusion. Here we consider a case where the symmetric method gives
negative temperature.

As before, consider a $2 \times 2$ grid with a hot zone ($T=10$) in
the first quadrant and cold temperature ($T=0.1$) in the rest; the
only difference from the previous test problem is that the magnetic
field lines are along the $x$- axis, $b_x=1$ and $b_y=0$ (see Figure
\ref{fig:fig5}). Reflective boundary conditions are used for temperature, 
as before.
Since there is no temperature gradient along the field lines for the
grid point $(i,j)$, we do not expect the temperature there to
change. While all other methods give a stationary temperature in
time, the symmetric method results in a heat flux out of the grid
$(i,j)$ through the corner at $(i-1/2,j+1/2)$.  With the initial
condition as shown in Figure \ref{fig:fig5}, the only non-vanishing
symmetric heat flux out of $(i,j)$ is, $q_{x,i-1/2,j+1/2}=- (9.9/2)
n\chi /\Delta x$. The only non-vanishing face-centered heat flux
entering the box through a face is $q_{x,i-1/2,j}=- (9.9/4) n\chi
/\Delta x<0$; i.e., heat flows out of $(i,j)$ which is already a
temperature minimum. This results in the temperature becoming
negative at $(i,j)$, although at late times it becomes equal to the
initial temperature at $(i,j)$. This simple test shows that the
symmetric method can also give negative temperatures (and associated
numerical problems) in presence of large temperature gradients.

\section{Slope limited fluxes}

As discussed earlier, the heat flux $q_x$ is composed of two terms:
the normal $q_{xx} = -n \chi b_x^2 \partial T/\partial x$ term, and
the transverse $q_{xy}=-n\chi b_xb_y \partial T/\partial y$ term.
For the asymmetric method the discrete form of the term $q_{xx} = -n
\chi b_x^2 \partial T/\partial x$ has the same sign as $-\partial
T/\partial x$, and hence guarantees that heat flows from higher to
lower temperatures. However, $q_{xy}=-n\chi b_xb_y
\partial T/\partial y$ can have an arbitrary sign, and can give rise
to heat flowing in the ``wrong" direction. We use slope limiters,
analogous to those used for linear reconstruction of variables in
numerical simulation of hyperbolic systems
\cite{VanLeer1979,Leveque2002}, to ``limit" the transverse terms.
Both asymmetric and symmetric methods can be modified with slope
limiters. The slope limited heat fluxes ensure that temperature
extrema are not accentuated. Thus, unlike the symmetric and
asymmetric methods, slope limited methods can never give negative
temperatures.

\subsection{Limiting the asymmetric method}

Since the normal heat flux term $q_{xx}$ is naturally located at
the face, no interpolation for $\partial T/\partial x$ is required
for its evaluation. However, an interpolation at the $x$- face is
required to evaluate $\overline{\partial T/\partial y}$ used in
$q_{xy}$ (the term with overlines in Eq. \ref{eq:q1_asymmetric}).
The arithmetic average used in Eq. (\ref{eq:asymmetric_Tavg}) for
$\ol{\partial T/\partial y}$ to calculate $q_{xy}$ was found to
result in heat flowing from lower to higher temperatures (see Figure
\ref{fig:fig4}). To remedy this problem we use slope limiters to
interpolate temperature gradients in the transverse heat flux term.

Slope limiters are widely used in numerical simulations of
hyperbolic equations (e.g., computational gas dynamics; see
\cite{VanLeer1979,Leveque2002}). Given the initial values for
 variables at grid centers, slope limiters (e.g., minmod, van Leer,
and Monotonized Central (MC)) are used to calculate the slopes of
conservative piecewise linear reconstructions in each grid cell.
Limiters use the variable values in the nearest grid cells to come
up with slopes that ensure that no new extrema are created for
the conserved variables along the characteristics, a property of
hyperbolic equations. Similarly, we use slope limiters to interpolate
temperature gradients in the transverse heat flux term so that unphysical
oscillations do not arise at temperature extrema.

The slope limited asymmetric heat flux in the $x$- direction is
still given by Eq. (\ref{eq:q1_asymmetric}), with the same
$\partial T/\partial x$ as in the asymmetric method, but a slope
limited interpolation for the transverse temperature gradient
$\overline{\partial T/\partial y}$, given by
\ba \nonumber \label{eq:slope_asymm} \left .
\overline{\frac{\partial T}{\partial y}} \right |_{i+1/2,j} &=& L
\left \{ L \left [\left . \frac{\partial T}{\partial y} \right
|_{i,j-1/2} , \left . \frac{\partial T}{\partial y} \right
|_{i,j+1/2} \right ], \right . \\
&& \hspace{1in} \left . L \left [ \left .
\frac{\partial T}{\partial y} \right |_{i+1,j-1/2}, \left .
\frac{\partial T}{\partial y} \right |_{i+1,j+1/2} \right ] \right
\}, \ea where $L$ is a slope limiter like minmod, van Leer, or
Monotonized Central (MC) limiter \cite{Leveque2002}; e.g., the MC 
limiter is given by 
\be
\label{eq:MC}
{\rm MC}(a,b) = {\rm minmod} \left [ 2~{\rm minmod}(a,b), \frac{a+b}{2} \right ],
\ee
where 
\ba
\nonumber
{\rm minmod}(a,b) &=& {\rm min}(a,b) \hspace{0.25in} {\rm if}~a, b > 0, \\
\nonumber
                  &=& {\rm max}(a,b) \hspace{0.25in} {\rm if}~a, b < 0, \\
\nonumber
                  &=& 0 \hspace{0.25in} {\rm if}~ab \leq 0.
\ea
A slope limiter weights the interpolation
towards the argument smallest in magnitude, if the arguments differ by too
much, and returns a zero if 
the two arguments are of opposite signs. An analogous expression for
the transverse temperature gradient at the $y$- face,
$\overline{\partial T/\partial x}$, is used to evaluate the heat
flux $q_y$. Interpolation similar to the asymmetric method is used for
all other variables (Eqs. \ref{eq:asymmetric_bavg} \&
\ref{eq:asymmetric_navg}).

\subsection{Limiting the symmetric method}

In the symmetric method, primary heat fluxes in both directions are
located at the cell corners (see Eq. \ref{eq:q1_symmetric}).
Temperature gradients in both directions have to be interpolated
at the corners. Thus, to ensure that temperature extrema are
not amplified with the symmetric method, both $\overline{\partial
T/\partial x}$ and $\overline{\partial T/\partial y}$ need to be
limited.

The face-centered $q_{xx,i+1/2,j}$ is calculated by averaging
$q_{xx}$ from the adjacent corners, which are given by the following
slope-limited expressions:
\ba \label{eq:qxx_symm_lim_up} q^N_{xx,i+1/2,j+1/2} &=&
-\overline{n}\overline{\chi} \overline{b_x^2} L2 \left [ \left .
\frac{\partial T}{\partial x} \right |_{i+1/2,j}, \left .
\frac{\partial T}{\partial x}
\right |_{i+1/2,j+1} \right ], \\
\label{eq:qxx_symm_lim_down} q^S_{xx,i+1/2,j-1/2} &=&
-\overline{n}\overline{\chi} \overline{b_x^2} L2 \left [ \left .
\frac{\partial T}{\partial x} \right |_{i+1/2,j}, \left .
\frac{\partial T}{\partial x} \right |_{i+1/2,j-1} \right ], \ea
where $N$ and $S$ superscripts indicate the north-biased and south-biased
heat fluxes. The face centered heat flux used in Eq.
(\ref{eq:e_evolve}) is
$q_{xx,i+1/2,j} = (q^N_{xx,i+1/2,j+1/2} + q^S_{xx,i+1/2,j-1/2})/2$;
the other interpolated quantities (indicated with an overline) are the
same as in Eq. (\ref{eq:q1_symmetric}). The limiter $L2$ which is
different from standard slope limiters is defined as
\ba \nonumber L2(a,b) &=& (a+b)/2, \mbox{ if } \min(\alpha a, a /
\alpha)
        < (a+b)/2 < \max(\alpha a, a /\alpha), \\
\nonumber
        &=&  \min(\alpha a, a / \alpha), \mbox{ if } (a+b)/2 \leq
             \min(\alpha a, a / \alpha), \\
        &=&  \max(\alpha a, a / \alpha), \mbox{ if } (a+b)/2 \geq
             \max(\alpha a, a / \alpha),
\ea
where $0<\alpha<1$ is a parameter; this reduces to a simple
averaging if the temperature is smooth, while restricting the
interpolated temperature ($\ol{\partial T/\partial x}$) to not
differ too much from $\partial T /
\partial x |_{i+1/2,j}$ (and be of the same sign).
We choose $\alpha=3/4$ for all of the results in this paper.
Note that the $L2$ limiter is not symmetric with
respect to its arguments. It ensures that $q_{xx,i+1/2,j \pm 1/2}$
is of the same sign as $-\partial T/\partial x |_{i+1/2,j}$; i.e.,
the interpolated normal heat flux is from higher to lower
temperatures. This interpolation will be able to diffuse the chess
board pattern in Figure \ref{fig:fig3}.  The transverse temperature 
gradient is limited in a way similar to the asymmetric method; the
temperature gradient $\ol{\partial T/\partial y}|_{i+1/2,j}$ is
still given by Eq. (\ref{eq:slope_asymm}).  Thus if $\alpha=1$, the
limited symmetric method becomes somewhat similar to the limited asymmetric
method (though with differences in the interpolation of the magnetic field
direction and of $n \chi$).

\section{Limiting with the entropy-like source function}
\label{sec:ent_limiting} If the entropy-like source function, which
we define as $\dot{s}^*=-\vec{q} \cdot \vec{\nabla} T$ (see Appendix
\ref{app:app5} to see how this is different from the entropy
function) is positive everywhere, heat is guaranteed to flow from
higher to lower temperatures. For the symmetric method, $\dot{s}^*$
evaluated at the cell corners is positive definite, but this need
not be true for interpolations at the cell faces; thus heat may flow
from lower to higher temperatures.
An entropy-like condition can be applied at all face-pairs to limit the
transverse heat flux terms ($q_{xy}$ and $q_{yx}$), such that
 \be \label{eq:ent_lim} \dot{s}^* = -
q_{x,i+1/2,j} \left . \frac{\partial T}{\partial x} \right
|_{i+1/2,j} - q_{y,i,j+1/2} \left . \frac{\partial T}{\partial y}
\right |_{i,j+1/2} \geq 0. \ee The limiter $L2$ is used to calculate
the normal gradients $q_{xx}$ and $q_{yy}$ at the faces, as in the
slope limited symmetric method (see \S 4.2). The use of $L2$ ensures
that $-q_{xx,i+1/2,j}\partial T/\partial x |_{i+1/2,j} \geq 0$, and
only the transverse terms $q_{xy}$ and $q_{yx}$ need to be reduced
to satisfy Eq. (\ref{eq:ent_lim}). That is, if on evaluating
$\dot{s}^*$ at all four face pairs the entropy-like condition (Eq.
\ref{eq:ent_lim}) is violated, the transverse terms are reduced to
make $\dot{s}^*$ vanish. The attractive feature of the entropy
limited symmetric method is that it reduces to the symmetric method
(which has the smallest numerical diffusion of all the methods; see
Figure \ref{fig:fig9}) when Eq. (\ref{eq:ent_lim}) is satisfied. The
hope is that limiting of transverse terms may prevent oscillations
with large temperature gradients.

The problem with entropy limiting, unlike the slope limited methods,
is that it does not guarantee that numerical oscillations at large
temperature gradients will be absent (e.g, see Figures
\ref{fig:fig4} and \ref{fig:fig7}). For example, when $\partial
T/\partial x|_{i+1/2,j} = \partial T/\partial y|_{i,j+1/2}=0$, Eq.
(\ref{eq:ent_lim}) is satisfied for arbitrary heat fluxes
$q_{x,i+1/2,j}$ and $q_{y,i,j+1/2}$. In such a case, transverse heat
fluxes $q_{xy}$ and $q_{yx}$ can cause heat to flow in the ``wrong"
direction, causing unphysical oscillations at temperature extrema.
However, this unphysical behavior occurs only for a few time steps,
after which the oscillations are damped.
The result is that the overshoots are not as pronounced and quickly
decay with time, unlike in the asymmetric and symmetric methods (see
Figures \ref{fig:fig6} \& \ref{fig:fig7}).
Although temperature extrema can be accentuated by the entropy
limited method, early on one can choose sufficiently small time
steps to ensure that temperature does not become negative; this is
equivalent to saying that the entropy limited method will not give
overshoots at late times (see Figure \ref{fig:fig7} and Tables
\ref{tab:tab1}-\ref{tab:tab4}). This trick will not work for the
centered symmetric and asymmetric methods where temperatures can be
negative even at late times (see Figure \ref{fig:fig7}).

To guarantee that temperature extrema are not amplified, in addition
to entropy limiting at all points, one can also use slope limiting
of transverse temperature gradients at extrema. This results in a
method that does not amplify the extrema, but is more diffusive
compared to just entropy limiting (see Figure \ref{fig:fig9}).
Because of the simplicity of slope limited methods and their
desirable mathematical properties (discussed in the next section),
they are preferred over the cumbersome entropy limited methods.

\section{Mathematical properties}
In this section we prove that the slope limited fluxes satisfy the
physical requirement that temperature extrema are not amplified.
Also discussed are global and local properties related to the
entropy-like condition $\dot{s}^* = - \vec{q} \cdot \vec{\nabla} T
\geq 0$.

\subsection{Behavior at temperature extrema}
Slope limiting of both asymmetric and symmetric methods guarantees
that temperature extrema are not amplified further, i.e., the
maximum temperature does not increase and the minimum temperature
does not decrease, as required physically. This ensures that the
temperature is always positive and numerical problems because of
imaginary sound speed do not arise. The normal heat flux in the
asymmetric method ($-\ol{n}\ol{\chi} b_x^2 \partial T/\partial x$)
and the $L2$ limited normal heat flux term in the symmetric method
(Eqs. \ref{eq:qxx_symm_lim_up} and \ref{eq:qxx_symm_lim_down})
allows the heat to flow only from higher to lower temperatures. Thus
the terms responsible for unphysical behavior at temperature extrema
are the transverse heat fluxes $q_{xy}$ and $q_{yx}$. Slope limiters
ensure that the transverse heat terms vanish at extrema and heat
flows down the temperature gradient at those grid points.

The operator $L(L(a,b),L(c,d))$, where $L$ is a slope limiter like
minmod, van Leer, or MC, is symmetric with respect to all its
arguments, and hence can be written as $L(a,b,c,d)$. For the slope
limiters considered here (minmod, van Leer, and MC), $L(a,b,c,d)$
vanishes unless all four arguments $a,b,c,d$ have the same sign. At
a local temperature extremum (say at $(i,j)$), the $x$- (and $y$-)
face-centered slopes $\partial T/\partial y|_{i,j+1/2}$ and
$\partial T/\partial y|_{i,j-1/2}$ (and $\partial T/\partial
x|_{i+1/2,j}$ and $\partial T/\partial x|_{i-1/2,j}$) are of
opposite signs, or at least one of them is zero. This ensures that
the slope limited transverse temperature gradients
($\overline{\partial T/\partial y}$ and $\overline{\partial
T/\partial x}$) vanish (from Eq. \ref{eq:slope_asymm}).
Thus, the heat fluxes are $q_{x,i \pm 1/2, j} = -\ol{n}\ol{\chi}
\ol{b_x}^2 \partial T/\partial x|_{i \pm 1/2, j}$ and $q_{y,i, j\pm
1/2} = -\ol{n}\ol{\chi} \ol{b_y}^2 \partial T/\partial y|_{i, j \pm
1/2}$ at the temperature extrema, which are always down the
temperature gradient. This ensures that temperature never decreases
(increases) at a temperature minimum (maximum), and negative
temperatures are avoided.

\subsection{The entropy-like condition, $\dot{s}^* = -\vec{q} \cdot
\vec{\nabla} T \geq 0$}

If the number density $n$ remains constant in time, then
multiplying Eq. (\ref{eq:anisotropic_conduction}) with $T$ and
integrating over all space gives
\ba \label{eq:selfadjointness}
\nonumber
\frac{1}{(\gamma-1)} \frac{\partial}{\partial t} \int n T^2 dV  = -
\int T \vec{\nabla} \cdot \vec{q} dV &=& \int \vec{q} \cdot \vec{\nabla}
T dV \\
&=& - \int n \chi |\nabla_\parallel T|^2 dV \le 0, \ea assuming that the
surface contributions vanish. This analytic constraint implies that
volume averaged temperature fluctuations cannot increase in time.
Locally it gives the entropy-like condition $\dot{s}^*=-\vec{q}
\cdot \vec{\grad} T \geq 0$, implying that heat always flows from higher
to lower temperatures.

Ref. \cite{Gunter2005} has shown that the symmetric method
satisfies  $\dot{s}^*=-\vec{q} \cdot \vec{\grad} T \geq 0$ at cell corners.
The entropy-like function $\dot{s}^*$ evaluated at $(i+1/2,j+1/2)$
with the symmetric method is  \be
\dot{s}^*_{i+1/2,j+1/2} =  -q_{x,i+1/2,j+1/2} \left.
\ol{\frac{\partial T}{\partial x}} \right |_{i+1/2,j+1/2} -
q_{y,i+1/2,j+1/2} \left. \ol{\frac{\partial T}{\partial y}} \right
|_{i+1/2,j+1/2}. \ee Using the symmetric heat fluxes (Eq.
\ref{eq:q1_symmetric}) the entropy-like function becomes, \ba
\nonumber \dot{s}^* &=& \ol{n}\ol{\chi} \ol{b_x}^2 \left | \ol{\frac{\partial
T}{\partial x}} \right |^2 + \ol{n}\ol{\chi} \ol{b_y}^2 \left |
\ol{\frac{\partial T}{\partial y}} \right |^2
+ 2 \ol{n}\ol{\chi} \ol{b_x}~\ol{b_y}
\ol{\frac{\partial T}{\partial x}}~\ol {\frac{\partial T}{\partial
y}}, \\ &=& \ol{n}\ol{\chi} \left [ \ol{b_x} \ol{\frac{\partial
T}{\partial x}} +  \ol{b_y} \ol{\frac{\partial T}{\partial y}}
\right ]^2 \geq 0, \ea
and integration over the whole space implies Eq. (\ref{eq:selfadjointness}).
Although the entropy-like condition is satisfied by the symmetric method
at grid corners (both locally and globally), this condition is not
sufficient to guarantee positivity of temperature at cell centers, as we
demonstrate in \S \ref{subsec:symmfail}.
Also notice that the modification of the symmetric method to satisfy the
entropy-like condition at face pairs (see \S \ref{sec:ent_limiting})
does not cure the problem of negative temperatures (see Figure \ref{fig:fig4}).
Thus, a method which satisfies the entropy-like condition
($\dot{s}^* = -\vec{q} \cdot \vec{\nabla} T \geq 0$) interpolated at
some point does not necessarily satisfy it everywhere, implying that
unphysical oscillations in the presence of large temperature gradients may
arise even if the interpolated entropy-like condition holds.

With an appropriate interpolation, the asymmetric method and the slope
limited asymmetric methods
can be modified to satisfy the global entropy-like condition $\dot{S}^* = -\int
\vec{q} \cdot \vec{\grad} T dV/V \geq 0$. Consider \be \dot{S}^* =
\frac{-1}{N_x N_y} \sum_{i,j} \left [ q_{x,i+1/2,j} \left .
\frac{\partial T}{\partial x} \right |_{i+1/2,j} + q_{y,i,j+1/2}
\left . \frac{\partial T}{\partial y} \right |_{i,j+1/2}  \right ],
\ee where $N_x$ and $N_y$ are the number of grid points in each
direction. Substituting the form of asymmetric heat fluxes,
\ba \nonumber \label{eq:Sdot}  \dot{S}^* &=& \frac{1}{N_x N_y}
\sum_{i,j} \left [ \left ( \ol{n} \ol{\chi} b_x^2 \left |
\frac{\partial T}{\partial x} \right |^2 \right )_{i+1/2,j} + \left
( \ol{n} \ol{\chi} b_y^2 \left | \frac{\partial T}{\partial
y} \right |^2 \right)_{i,j+1/2} \right . \\
&+& \left . \left ( \ol{n \chi b_x b_y \frac{\partial T}{\partial
y}} \right )_{i+1/2,j} \left . \frac{\partial T}{\partial x} \right
|_{i+1/2,j} +  \left ( \ol{n \chi b_x b_y \frac{\partial T}{\partial
x}} \right )_{i,j+1/2} \left . \frac{\partial T}{\partial y} \right
|_{i,j+1/2} \right ], \ea where overlines represent appropriate
interpolations.
We define \ba G_{x,i+1/2,j} &=& \sqrt{ \left( \ol{n}\ol{\chi}
\right)}_{i+1/2,j} b_{x,i+1/2,j} \left . \frac{\partial
T}{\partial x} \right |_{i+1/2,j}, \\
G_{y,i,j+1/2} &=& \sqrt{ \left( \ol{n}\ol{\chi} \right)}_{i,j+1/2}
b_{y,i,j+1/2} \left . \frac{\partial
T}{\partial y} \right |_{i,j+1/2}, \\
\ol{G}_{y,i+1/2,j} &=& \ol{ \sqrt{n \chi}
b_y \left . \frac{\partial
T}{\partial y} \right | }_{i+1/2,j}, \\
\ol{G}_{x,i,j+1/2} &=& \ol{ \sqrt{n \chi}
b_x \left . \frac{\partial T}{\partial x} \right |
}_{i,j+1/2}.
 \ea
In terms of $G$'s, Eq. (\ref{eq:Sdot}) can be written as
\ba \nonumber \dot{S}^*
= \frac{1}{N_x N_y} \sum_{i,j} && \left [  G_{x,i+1/2,j}^2 +
G_{y,i,j+1/2}^2 \right . \\
&+& \left . G_{x,i+1/2,j} \ol{G}_{y,i+1/2,j} +
\ol{G}_{x,i,j+1/2} G_{y,i,j+1/2} \right ]. \ea

A lower bound on $\dot{S}^*$ is obtained by assuming the cross terms
to be negative, i.e.,
\ba \nonumber \dot{S}^* \geq  \frac{1}{N_x N_y} \sum_{i,j} &&
\left [ G_{x,i+1/2,j}^2 + G_{y,i,j+1/2}^2 \right . \\
&-& \left . \left | G_{x,i+1/2,j}
 \ol{G}_{y,i+1/2,j} \right | - \left |
\ol{G}_{x,i,j+1/2} G_{y,i,j+1/2} \right | \right ]. \ea Now define
$\ol{G}_{y,i+1/2,j}$ and $\ol{G}_{x,i,j+1/2}$ as follows (the
following interpolation is necessary for the proof to hold): \ba
\ol{G}_{x,i,j+1/2} &=& L ( G_{x,i+1/2,j}, G_{x,i-1/2,j},
G_{x,i+1/2,j+1}, G_{x,i-1/2,j+1} ), \\
\ol{G}_{y,i+1/2,j} &=& L ( G_{y,i,j+1/2}, G_{y,i,j-1/2},
G_{y,i+1,j+1/2}, G_{y,i+1,j-1/2} ),
\ea where $L$ is an arithmetic
average (as in centered asymmetric method) or a slope limiter
(e.g., minmod, van Leer, or MC) which satisfies the property that $
|L(a,b,c,d)| \leq (|a|+|b|+|c|+|d|)/4$.
Thus, \ba \nonumber \dot{S}^* &\ge& \frac{1}{N_xN_y}
\sum_{i,j} G_{x,i+1/2,j}^2 + G_{y,i,j+1/2}^2  - \frac{1}{4} \left [
\left |G_{x,i+1/2,j} G_{y,i,j+1/2} \right | \right . \\ \nonumber
&+& \left |G_{x,i+1/2,j} G_{y,i,j-1/2} \right |  + \left
|G_{x,i+1/2,j} G_{y,i+1,j+1/2} \right | + \left |G_{x,i+1/2,j}
G_{y,i+1,j-1/2} \right |
\\  \nonumber &+& \left |G_{y,i,j+1/2} G_{x,i+1/2,j}\right | + \left |
G_{y,i,j+1/2} G_{x,i-1/2,j} \right | + \left |G_{y,i,j+1/2} G_{x,i+1/2,j+1}
\right |  \\
&+& \left . \left |G_{y,i,j+1/2} G_{x,i-1/2,j+1} \right | \right ].
\ea Shifting the dummy indices and combining various terms give, \ba
\nonumber \dot{S}^* &\ge& \frac{1}{N_xN_y} \sum_{i,j} G_{x,i+1/2,j}^2
+ G_{y,i,j+1/2}^2 -\frac{1}{2} \left [  \left |G_{x,i+1/2,j}
G_{y,i,j+1/2} \right |  \right . \\
&+& \nonumber \left . \left |G_{x,i+1/2,j} G_{y,i,j-1/2} \right | +
\left | G_{x,i+1/2,j} G_{y,i+1,j+1/2} \right |
+ \left |G_{x,i+1/2,j} G_{y,i+1,j-1/2} \right | \right ] \\
\nonumber &=& \frac{1}{4N_xN_y} \sum_{i,j} \left [ \left (
|G_{x,i+1/2,j}| - |G_{y,i,j+1/2}| \right )^2 + \left (
|G_{x,i+1/2,j}| - |G_{y,i,j-1/2}| \right )^2 \right . \\
\nonumber &+& \left . \left ( |G_{x,i+1/2,j}| - |G_{y,i+1,j+1/2}|
\right )^2  + \left ( |G_{x,i+1/2,j}| - |G_{y,i+1,j-1/2}|
\right )^2 \right ] \\
&\geq & 0. \ea Thus, an appropriate interpolation for the asymmetric
and the slope limited asymmetric methods results in a scheme that
satisfies the global entropy-like condition. A variation of this
proof can be used to prove the global entropy condition $\dot{S}
\geq 0$ by multiplying Eq. (\ref{eq:anisotropic_conduction}) with
$1/T$ instead of $T$ (see Appendix \ref{app:app5}), although the
form of interpolation would need to be modified slightly. It is
comforting that introducing a limiter to the asymmetric method does
not break the global entropy-like condition. However, it is
important to remember that the entropy-like (or entropy) condition
satisfied at some point does not guarantee a local heat flow in the
correct direction; thus it is necessary to use slope limiters at
temperature extrema to avoid temperature oscillations.

\section{Further tests}
\begin{sidewaystable}
\centering
\begin{tabular}{ccccccc}
\hline
Method   & L1 error & L2 error  & L$\infty$ error & T$_{\rm max}$ & T$_{\rm min}$ &
$\chi_{\perp,{\rm num}}/\chi_\parallel$ \\
\hline
asymmetric & 0.0324 & 0.0459 & 0.0995 & 10.0926 & 9.9744 & 0.0077 \\
asymmetric minmod & 0.0471 & 0.0627 & 0.1195 & 10.0410 & 10 & 0.0486 \\
asymmetric MC & 0.0358 & 0.0509 & 0.1051 & 10.0708 & 10 & 0.0127 \\
asymmetric van Leer & 0.0426 & 0.0574 & 0.1194 & 10.0519 & 10 & 0.0238 \\
symmetric & 0.0114 & 0.0252 & 0.1425 & 10.2190 & 9.9544 & 0.00028 \\
symmetric entropy & 0.0333 & 0.0477 & 0.0997 & 10.0754 & 10 & 0.0088 \\
symmetric entropy extrema & 0.0341 & 0.0487 & 0.1010 & 10.0751 & 10 & 0.0101 \\
symmetric minmod & 0.0475 & 0.0629 & 0.1322 & 10.0406 & 10 & 0.0490 \\
symmetric MC & 0.0289 & 0.0453 & 0.0872 & 10.0888 & 10 & 0.0072 \\
symmetric van Leer & 0.0438 & 0.0585 & 0.1228 & 10.0519 & 10 & 0.0238 \\
\hline
\end{tabular}
\caption{Diffusion in circular field lines: $50 \times 50$ grid \label{tab:tab1} }
The errors are based on the assumption that the initial hot patch
has diffused to a uniform temperature ($T=10.1667$) in the ring 0.5$<r<$0.7,
and $T=10$ outside it.
\end{sidewaystable}

\begin{sidewaystable}
\centering
\begin{tabular}{ccccccc}
\hline
Method   & L1 error & L2 error  & L$\infty$ error & T$_{\rm max}$ & T$_{\rm min}$ &
$\chi_{\perp, {\rm num}}/\chi_\parallel$ \\
\hline
asymmetric & 0.0256 & 0.0372 & 0.0962 & 10.1240 & 9.9859 & 0.0030 \\
asymmetric minmod & 0.0468 & 0.0616 & 0.1267 & 10.0439 & 10 & 0.0306 \\
asymmetric MC & 0.0261 & 0.0405 & 0.0907 & 10.1029 & 10 & 0.0040 \\
asymmetric van Leer & 0.0358 & 0.0502 & 0.1002 & 10.0741 & 10 & 0.0971 \\
symmetric & 0.0079 & 0.0173 & 0.1206 & 10.2276 & 9.9499 & $4.1 \times 10^{-5}$ \\
symmetric entropy & 0.0285 & 0.0420 & 0.0881 & 10.0961 & 10 & 0.0042 \\
symmetric entropy extrema & 0.0291 & 0.0425 & 0.0933 & 10.0941 & 10 & 0.0041 \\
symmetric minmod & 0.0471 & 0.0618 & 0.1275 & 10.0433 & 10 & 0.0305 \\
symmetric MC & 0.0123 & 0.0252 & 0.1133 & 10.1406 & 10 & 0.00084 \\
symmetric van Leer & 0.0374 & 0.0514 & 0.1038 & 10.0697 & 10 & 0.0104 \\
\hline
\end{tabular}
\caption{Diffusion in circular field lines: $100 \times 100$ grid \label{tab:tab2} }
\end{sidewaystable}

\begin{sidewaystable}
\centering
\begin{tabular}{ccccccc}
\hline
Method   & L1 error & L2 error  & L$\infty$ error & T$_{\rm max}$ & T$_{\rm min}$ &
$\chi_{\perp, {\rm num}}/\chi_\parallel$ \\
\hline
asymmetric & 0.0165 & 0.0281 & 0.0949 & 10.1565 & 9.9878 & 0.0012 \\
asymmetric minmod & 0.0441 & 0.0585 & 0.1214 & 10.0511 & 10 & 0.0191 \\
asymmetric MC & 0.0161 & 0.0289 & 0.0930 & 10.1397 & 10 & 0.0015 \\
asymmetric van Leer & 0.0264 & 0.0407 & 0.0928 & 10.1006 & 10 & 0.0035 \\
symmetric & 0.0052 & 0.0132 & 0.1125 & 10.2216 & 9.9509 & $1.9 \times 10^{-5}$  \\
symmetric entropy & 0.0256 & 0.0385 & 0.0959 & 10.1103 & 10 & 0.0032 \\
symmetric entropy extrema & 0.0260 & 0.0391 & 0.0954 & 10.1074 & 10 & 0.0032 \\
symmetric minmod & 0.0444 & 0.0588 & 0.1219 & 10.0503 & 10 & 0.0192 \\
symmetric MC & 0.0053 & 0.0160 & 0.0895 & 10.1676 & 10 & 0.0002 \\
symmetric van Leer & 0.0281 & 0.0426 & 0.0901 & 10.0952 & 10 & 0.0038 \\
\hline
\end{tabular}
\caption{Diffusion in circular field lines: $200 \times 200$ grid \label{tab:tab3} }
\end{sidewaystable}

\begin{sidewaystable}
\centering
\begin{tabular}{ccccccc}
\hline
Method   & L1 error & L2 error  & L$\infty$ error & T$_{\rm max}$ & T$_{\rm min}$ &
$\chi_{\perp, {\rm num}}/\chi_\parallel$ \\
\hline
asymmetric & 0.0118 & 0.0234 & 0.0866 & 10.1810 & 9.9898 & $5.9 \times 10^{-4}$ \\
asymmetric minmod & 0.0399 & 0.0539 & 0.1120 & 10.0629 & 10 & 0.0115 \\
asymmetric MC & 0.0102 & 0.0230 & 0.0894 & 10.1708 & 10 &  $6.8 \times 10^{-4}$ \\
asymmetric van Leer & 0.0167 & 0.0290 & 0.1000 & 10.1321 & 10 & 0.0013 \\
symmetric & 0.0033 & 0.0104 & 0.1112 & 10.2196 & 9.9504 & $8.4 \times 10^{-6}$  \\
symmetric entropy & 0.0252 & 0.0384 & 0.0969 & 10.1144 & 10 & 0.0027 \\
symmetric entropy extrema & 0.0253 & 0.0383 & 0.0958 & 10.1135 & 10 & 0.0026 \\
symmetric minmod & 0.0401 & 0.0541 & 0.1124 & 10.0622 & 10 & 0.0116 \\
symmetric MC & 0.0032 & 0.0122 & 0.0896 & 10.1698 & 10 & $6.5 \times 10^{-5} $ \\
symmetric van Leer & 0.0182 & 0.0307 & 0.1026 & 10.1260 & 10 & 0.0013 \\
\hline
\end{tabular}
\caption{Diffusion in circular field lines: $400 \times 400$ grid \label{tab:tab4} }
\end{sidewaystable}
We use test problems discussed in \cite{Parrish2005} and
\cite{Sovinec2004} to compare different methods. The first test
problem (taken from \cite{Parrish2005}) initializes a hot patch in
circular field lines; ideally the hot patch should diffuse only
along the field lines, but perpendicular numerical diffusion causes
some cross-field thermal conduction. Unlike the limited methods,
both asymmetric and symmetric methods show temperature oscillations
at the temperature discontinuity. The second test problem (from
\cite{Sovinec2004}) includes a source term and an explicit
perpendicular diffusion coefficient ($\chi_\perp$). The steady state
temperature gives a measure of the perpendicular numerical diffusion
$\chi_{\perp,{\rm num}}$.

\subsection{Diffusion of a hot patch in circular magnetic field}
\begin{figure}
\centering
\includegraphics[width=2.5in,height=2.0in]{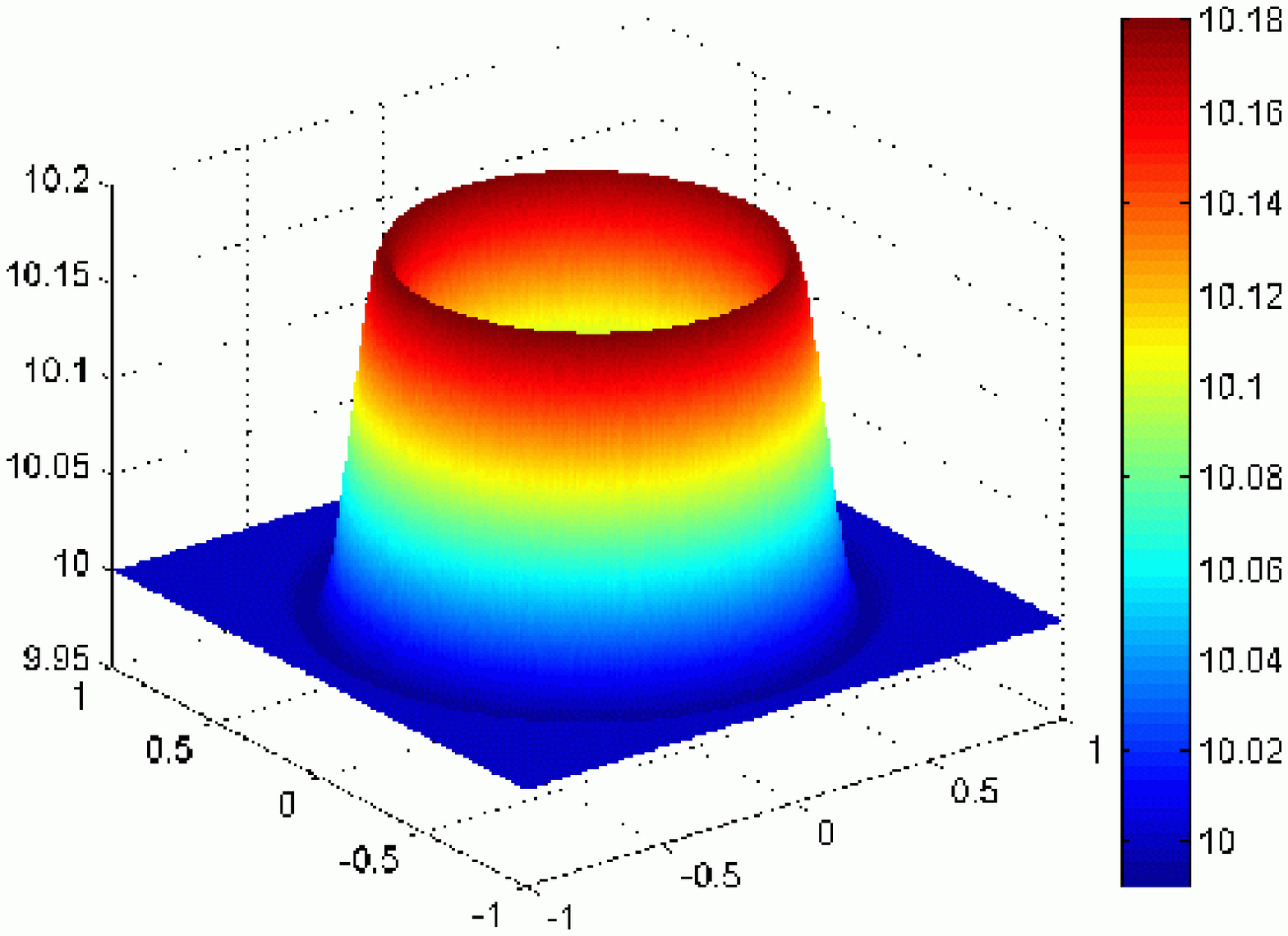}
\includegraphics[width=2.5in,height=2.0in]{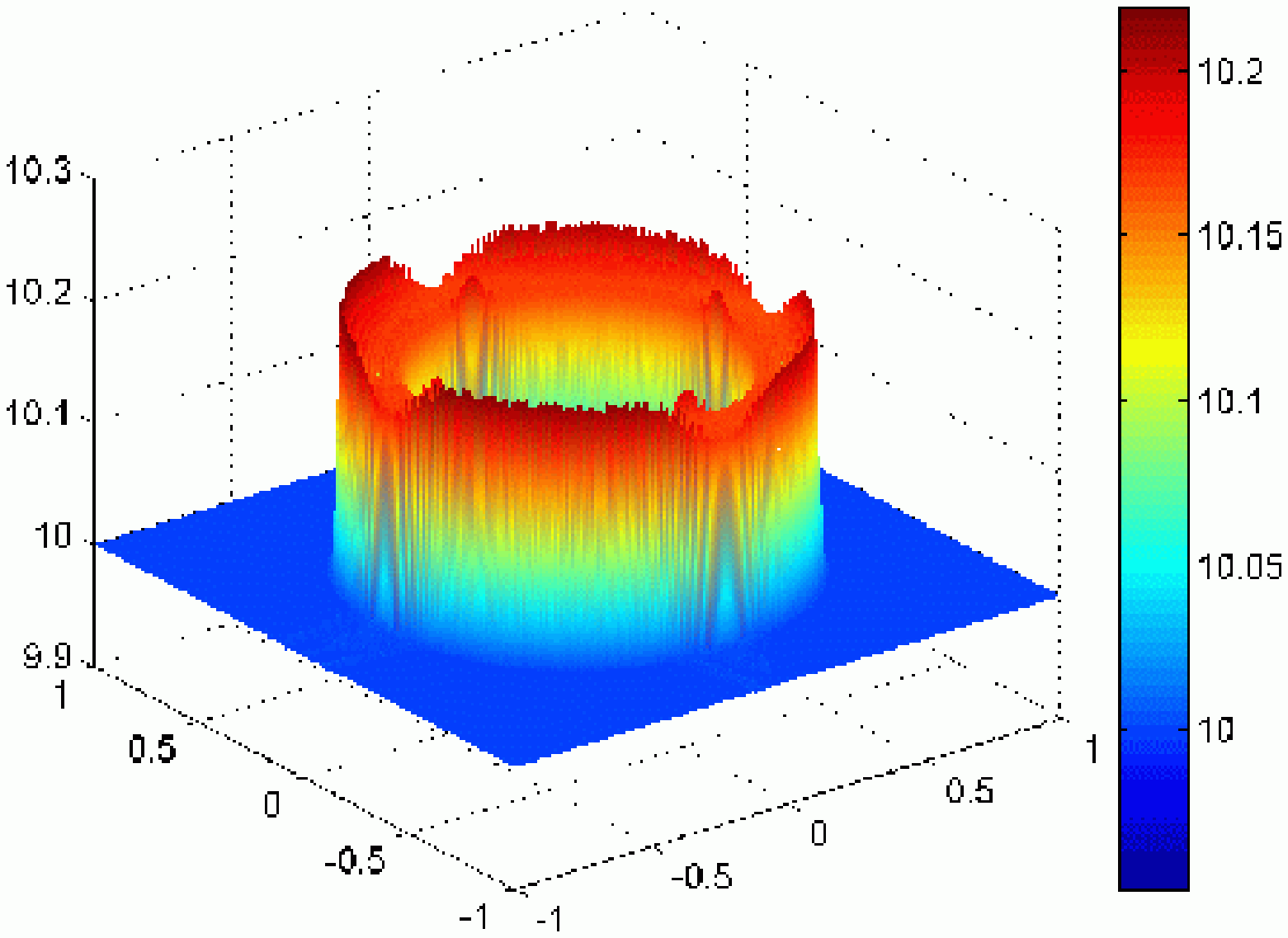}
\includegraphics[width=2.5in,height=2.0in]{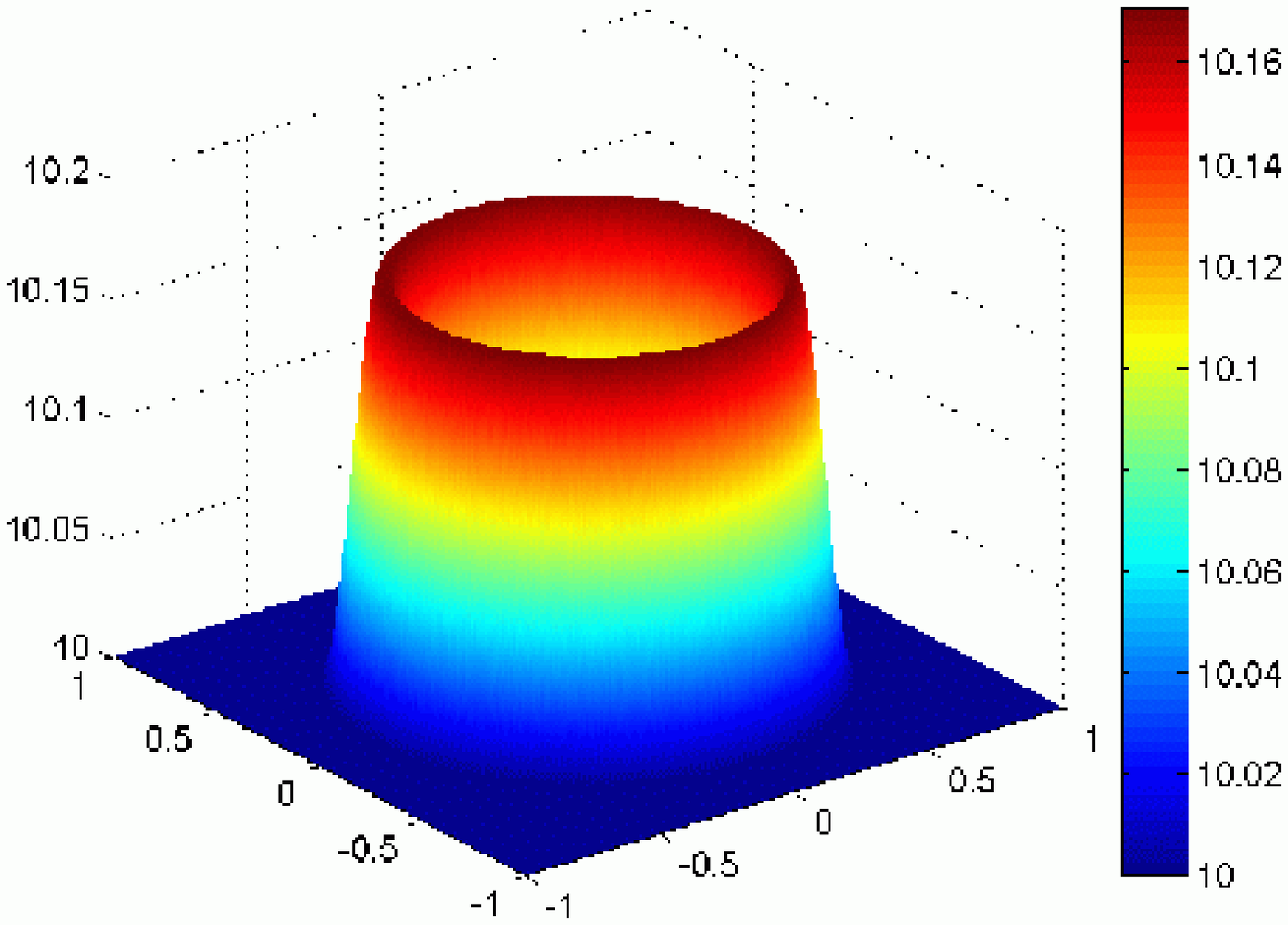}
\includegraphics[width=2.5in,height=2.0in]{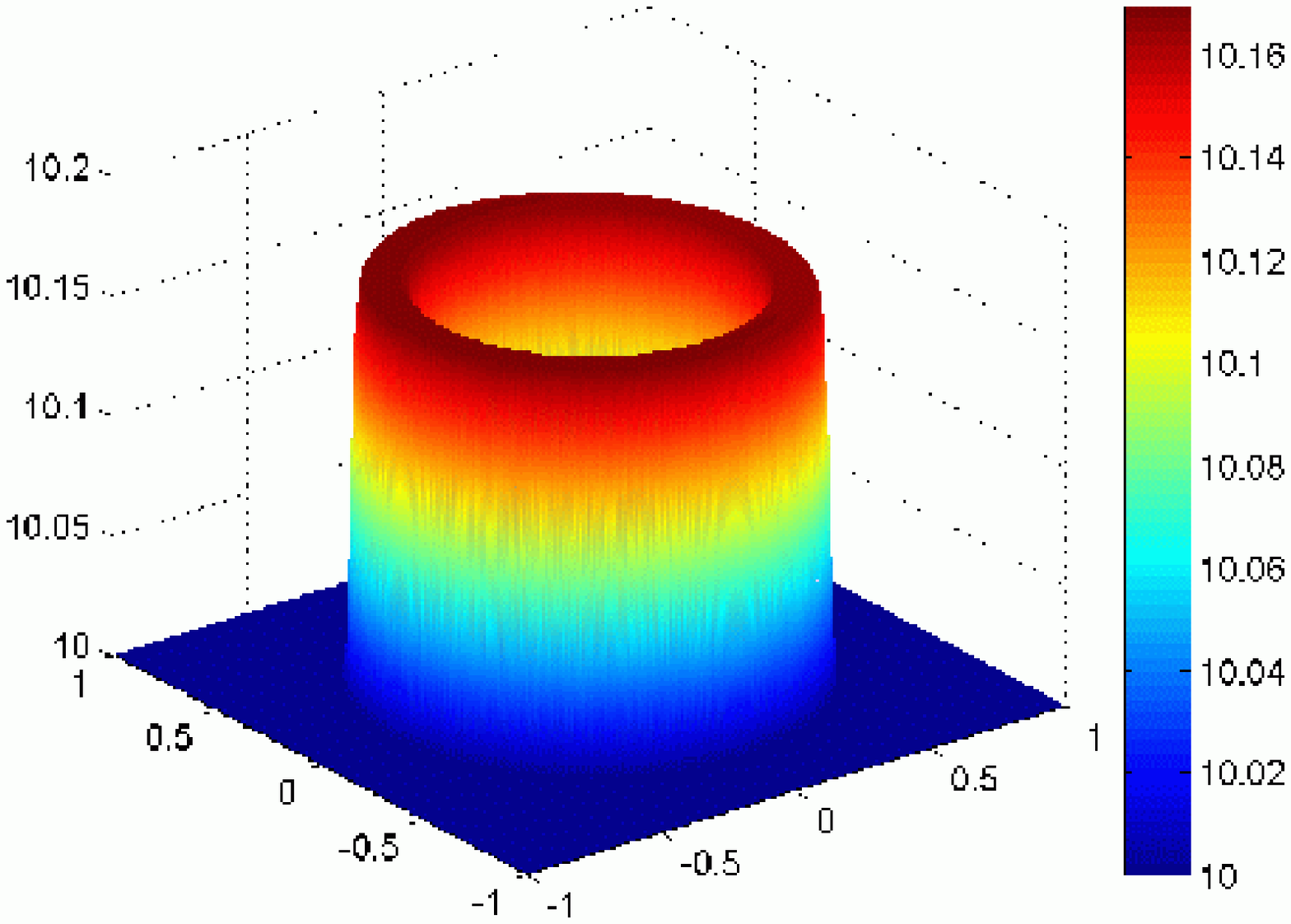}
\includegraphics[width=2.5in,height=2.0in]{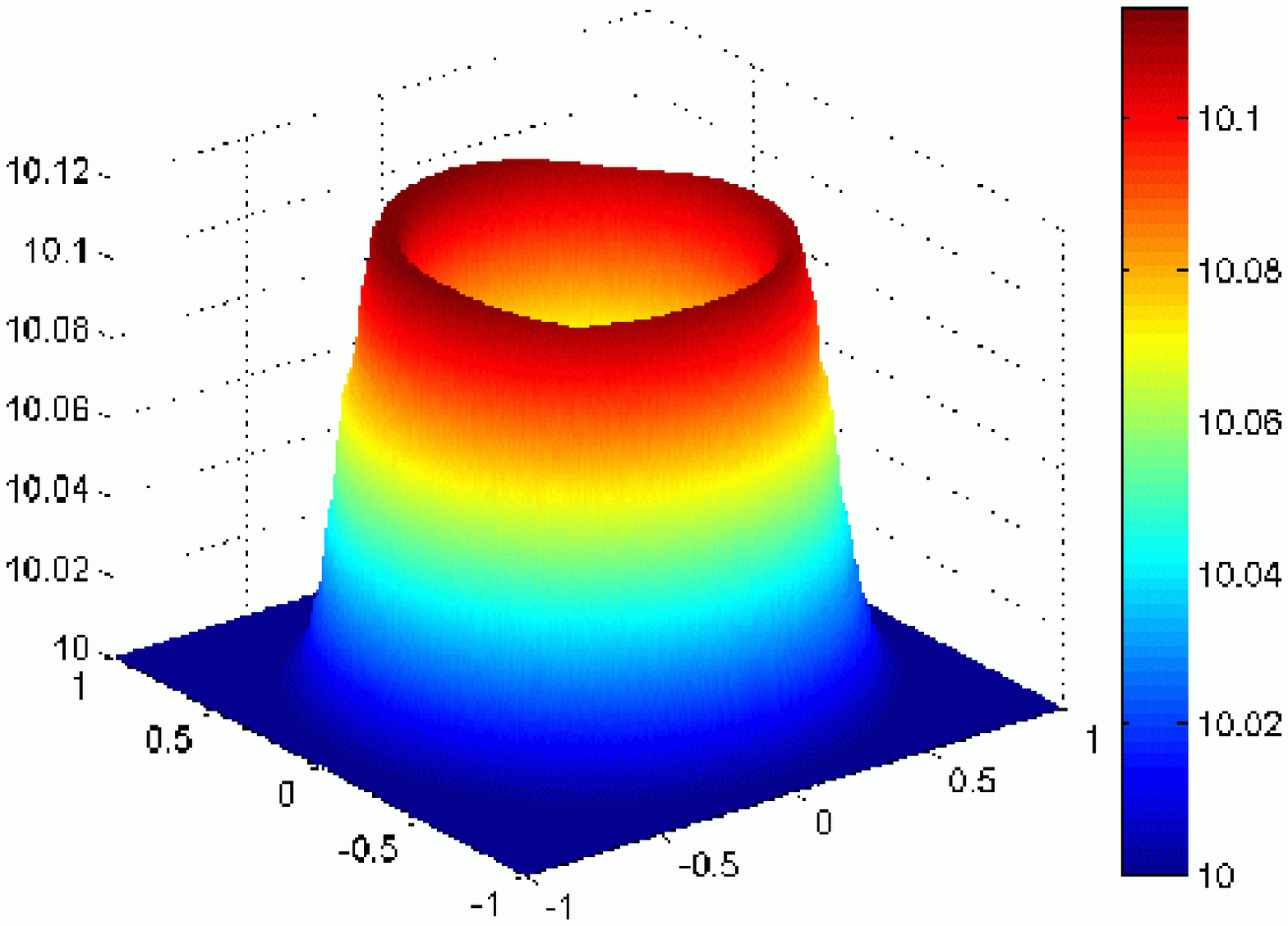}
\includegraphics[width=2.5in,height=2.0in]{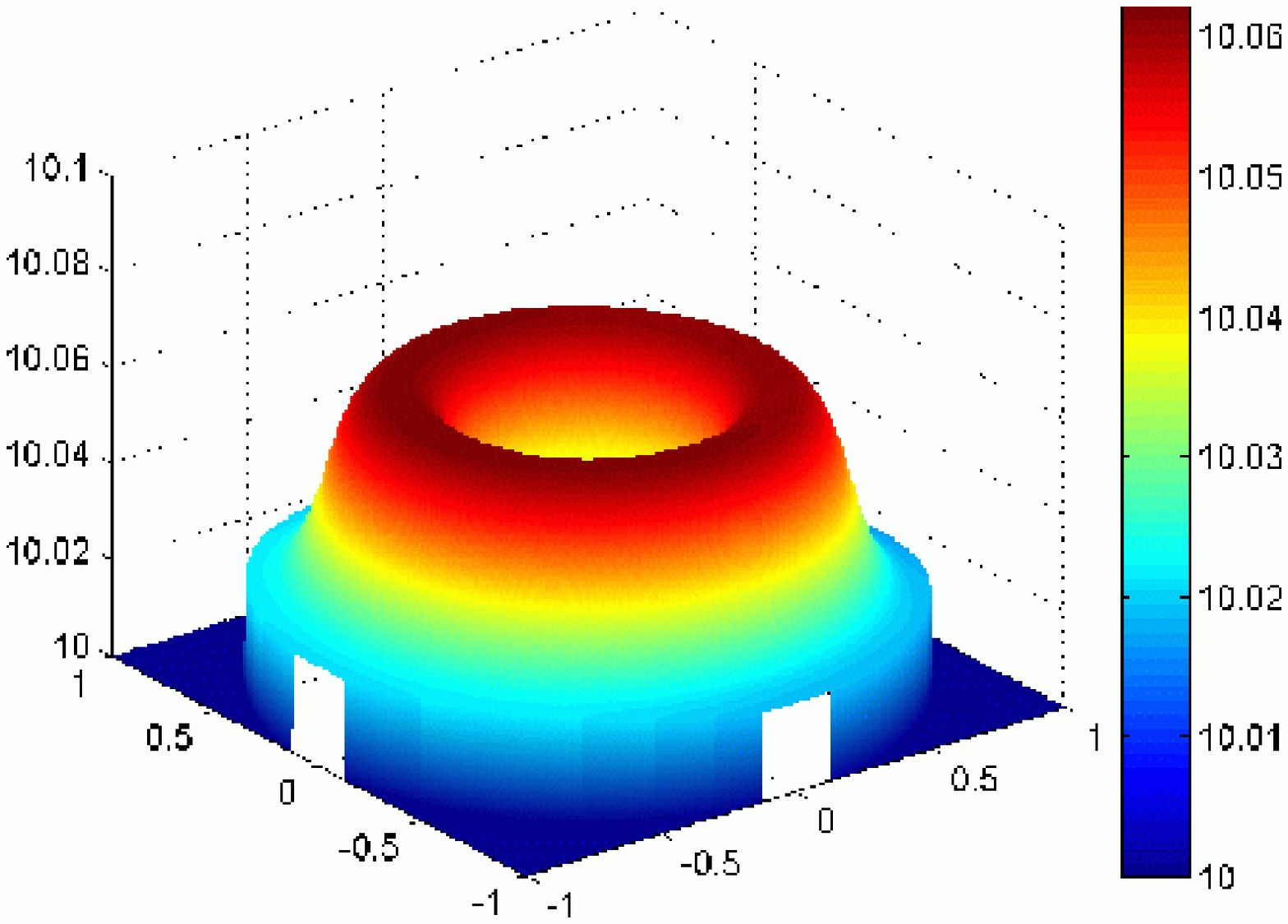}
\caption{The temperature at $t=200$ for
different methods initialized with the ring diffusion problem on a
400 $\times$ 400 grid. Shown from left to right and top to bottom
are the temperatures for: asymmetric, symmetric, asymmetric-MC,
symmetric-MC, entropy limited symmetric, and minmod methods. Both
asymmetric and symmetric methods give temperatures below 10 (the
initial minimum temperature) at late times. The result with a
minmod limiter is
very diffusive. The slope limited symmetric method is less diffusive
than the slope limited asymmetric method. Entropy limited method
does not show non-monotonic behavior at late times, but is diffusive
compared to the better slope limited methods.\label{fig:fig6}}
\end{figure}

The circular diffusion test problem was proposed in
\cite{Parrish2005}. A hot patch surrounded by a cooler background is
initialized in circular field lines; the temperature drops
discontinuously across the patch boundary. At late times, we expect
the temperature to become uniform (and higher) in a ring along the
magnetic field lines. The computational domain is a
$[-1,1]\times[-1,1]$ Cartesian box. The initial temperature distribution 
is given by \ba
\label{eq:ring_problem} \nonumber T &=& 12 \hspace{0.25 in}
\mbox{if} \hspace{0.1 in} 0.5<r<0.7 \hspace{0.1 in}
\mbox{and} \hspace {0.1 in}  \frac{11}{12}\pi<\theta<\frac{13}{12}\pi, \\
&=& 10 \hspace{0.25 in} \mbox{otherwise}, \ea where
$r=\sqrt{x^2+y^2}$ and $\tan\theta=y/x$. Fixed circular magnetic
field lines centered at the origin are initialized and number density ($n$) 
is set to unity. Reflective boundary conditions are
used for temperature; magnetic field and conduction vanishes outside $r=1$.
The parallel conduction coefficient $\chi=0.01$; there is no explicit
perpendicular diffusion ($\chi_\perp=0$). We evolve the anisotropic
conduction equation (\ref{eq:e_evolve}) till time = 200, by when we
expect the temperature to be almost uniform along the circular ring
$0.5<r<0.7$. In steady state (at late times), energy conservation
implies that the ring temperature should be 10.1667, while the
temperature outside the ring should be maintained at 10.

Figure \ref{fig:fig6} shows the temperature distribution for
different methods at time=200. All methods result in a higher
temperature in the annulus $r \in [0.5,0.7]$. The limited
schemes show larger perpendicular diffusion (see Tables
\ref{tab:tab1}-\ref{tab:tab4} which give errors, minimum and maximum
temperatures, and numerical perpendicular diffusion at time=200; also
see Figure \ref{fig:fig8})
compared to the symmetric and asymmetric schemes. The perpendicular
numerical diffusion ($\chi_{\perp,{\rm num}}$) scales with the parallel
diffusion coefficient $\chi$ for all methods. Notice that for Sovinec's
test problem (discussed in the next section) where temperature is
smooth and an explicit $\chi_\perp$ is present, perpendicular numerical
diffusion for the symmetric method does not increase with increasing
$\chi_\parallel$.

\begin{figure}
\centering
\includegraphics[width=4in,height=3in]{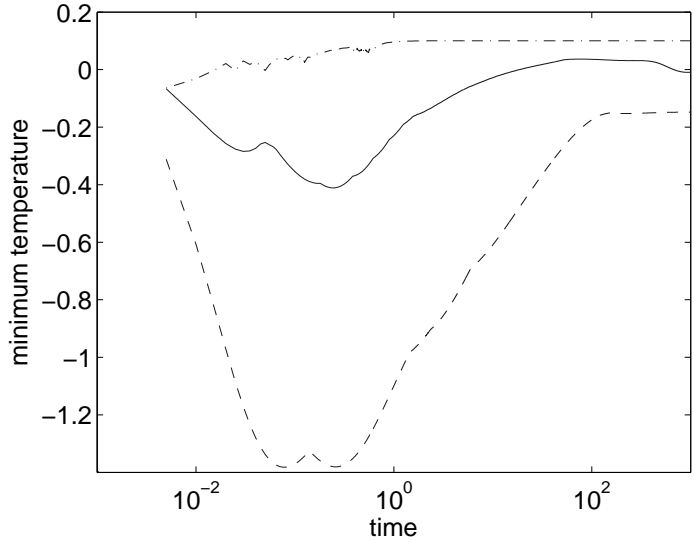}
\caption{Minimum temperature over the whole box as a function of time
for the ring diffusion test problem: symmetric (dashed line),
asymmetric (solid line), and entropy limited symmetric
(dot dashed line) methods are shown.
Initially the temperature of the hot patch is 10 and the
background is at 0.1. Both asymmetric and symmetric methods result in
negative temperature, even at late times. The non-monotonic behavior
with the entropy limited method is considerably less pronounced; the
minimum temperature quickly becomes equal to the initial minimum
$0.1$. The slope limited heat fluxes maintain the minimum temperature
at 0.1 at all times, as required physically.\label{fig:fig7}}
\end{figure}

\begin{figure}
\centering
\includegraphics[width=4in,height=3in]{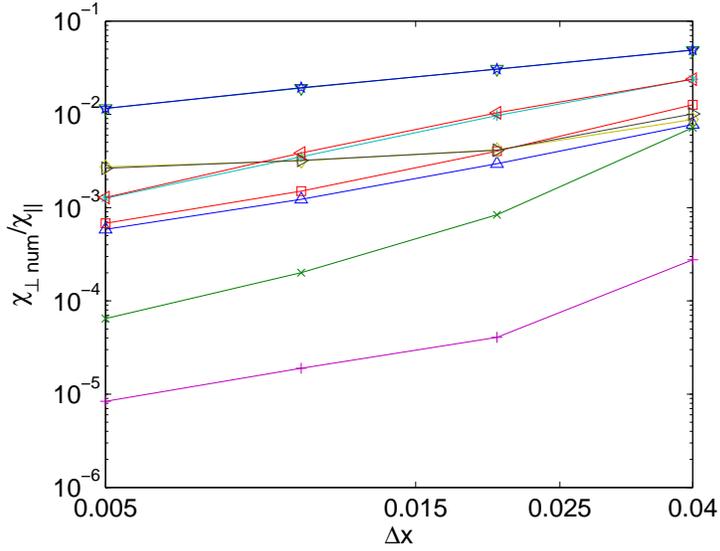}
\caption{Convergence of
$\chi_{\perp,{\rm num}}/\chi_\parallel$ as the number of grid points is
increased for the ring diffusion problem. The numerical
perpendicular diffusion $\chi_{\perp,{\rm num}}$ is calculated numerically
by measuring the heat diffusing out of the circular ring
(Eq. \ref{eq:chiperp_num}). The different
schemes are: asymmetric ($\triangle$), asymmetric with minmod
($\triangledown$), asymmetric with MC ($\square$), asymmetric with
van Leer ($\ast$), symmetric ($+$), symmetric with entropy limiting
($\diamond$), symmetric with entropy and extrema limiting
($\triangleright$), symmetric with minmod ($\star$), symmetric with
MC ($\times$), and symmetric with van Leer limiter
($\triangleleft$).
\label{fig:fig8}}
\end{figure}

The minmod limiter is much more diffusive than van Leer and MC
limiters. Both symmetric and asymmetric methods give a minimum
temperature below the initial minimum of 10, even at late times.
At late times the
symmetric method gives a temperature profile full of non-monotonic
oscillations (Figure \ref{fig:fig6}). Although the slope limited
fluxes are more diffusive than the symmetric and asymmetric methods,
they never show undershoots below the minimum temperature. The entropy
limited symmetric method gives temperature undershoots at early times
which are damped quickly, and the minimum temperature is still $10$
at late times (see Tables \ref{tab:tab1}-\ref{tab:tab4} \&
Figure \ref{fig:fig7}). Entropy limiting combined with a slope
limiter at temperature extrema behaves similar to the slope limiter
based schemes.

Strictly speaking, a hot ring surrounded by a cold background is not
a steady solution for the ring diffusion problem. Temperature in the
ring will diffuse in the perpendicular direction (because of
perpendicular numerical diffusion, although very slowly) until the
whole box is at a constant temperature. A rough estimate for time
averaged perpendicular numerical diffusion $\la \chi_{\perp,{\rm num}}
\ra$ follows from Eq. (\ref{eq:anisotropic_conduction}), \be
\label{eq:chiperp_num} \la \chi_{\perp,{\rm num}} \ra = \frac{ \int
(T_f - T_i) dV} {\int dt \left ( \int \grad^2 T dV \right )}, \ee
where the space integral is taken over the hot ring $0.5<r<0.7$, and
$T_i$ and $T_f$ are the initial and final temperature distributions
in the ring. Figure \ref{fig:fig8} plots the numerical
perpendicular diffusion (using Eq. \ref{eq:chiperp_num}) for the
ring diffusion problem at different resolutions (see Tables
\ref{tab:tab1}-\ref{tab:tab4}). The estimates
for perpendicular diffusion agree roughly with the more accurate
calculations using Sovinec's test problem described in the next
section (compare Figures \ref{fig:fig8} \&
\ref{fig:fig9}); as with Sovinec's test, the symmetric method is
the least diffusive. Table \ref{tab:tab5} lists the convergence rate
of $\chi_{\perp,{\rm num}}$ for the ring diffusion problem evolved with
different methods.
\begin{table}[hbt]
\centering
\caption{Asymptotic slopes for convergence of $\chi_{\perp, {\rm num}}$ in
the ring diffusion test problem \label{tab:tab5}}
\begin{tabular}{cc}
\hline
Method & slope \\
\hline
asymmetric & 1.066  \\
asymmetric minmod & 0.741 \\
asymmetric MC & 1.142 \\
asymmetric van Leer & 1.479 \\
symmetric & 1.181 \\
symmetric entropy & 0.220 \\
symmetric entropy extrema & 0.282 \\
symmetric minmod & 0.735 \\
symmetric MC & 1.636 \\
symmetric van Leer & 1.587 \\
\hline
\end{tabular}
\end{table}

To study the very long time behavior of different methods (in
particular to check whether the symmetric and asymmetric methods
give negative temperatures even at very late times) we initialize
the same problem with the hot patch at 10 and the cooler background
at 0.1. Figure \ref{fig:fig7} shows the minimum temperature with
time for the symmetric, asymmetric, and entropy limited symmetric
methods; slope limited methods give the correct result for the
minimum temperature ($T_{\rm min}=0.1$) at all times. With a large
temperature contrast, both symmetric and asymmetric methods give
negative minimum temperature even at late times.
Such points where temperature becomes negative, when coupled with
MHD equations, can give numerical instabilities because of an
imaginary sound speed.
The minimum temperature with the entropy limited symmetric method
shows small undershoots at early times which are damped quickly and
the minimum temperature is
equal to the initial minimum ($0.1$) after time=1.

\subsection{Convergence studies: measuring $\chi_{\perp, {\rm num}}$}
\begin{figure}
\centering
\includegraphics[width=4 in, height=3 in]{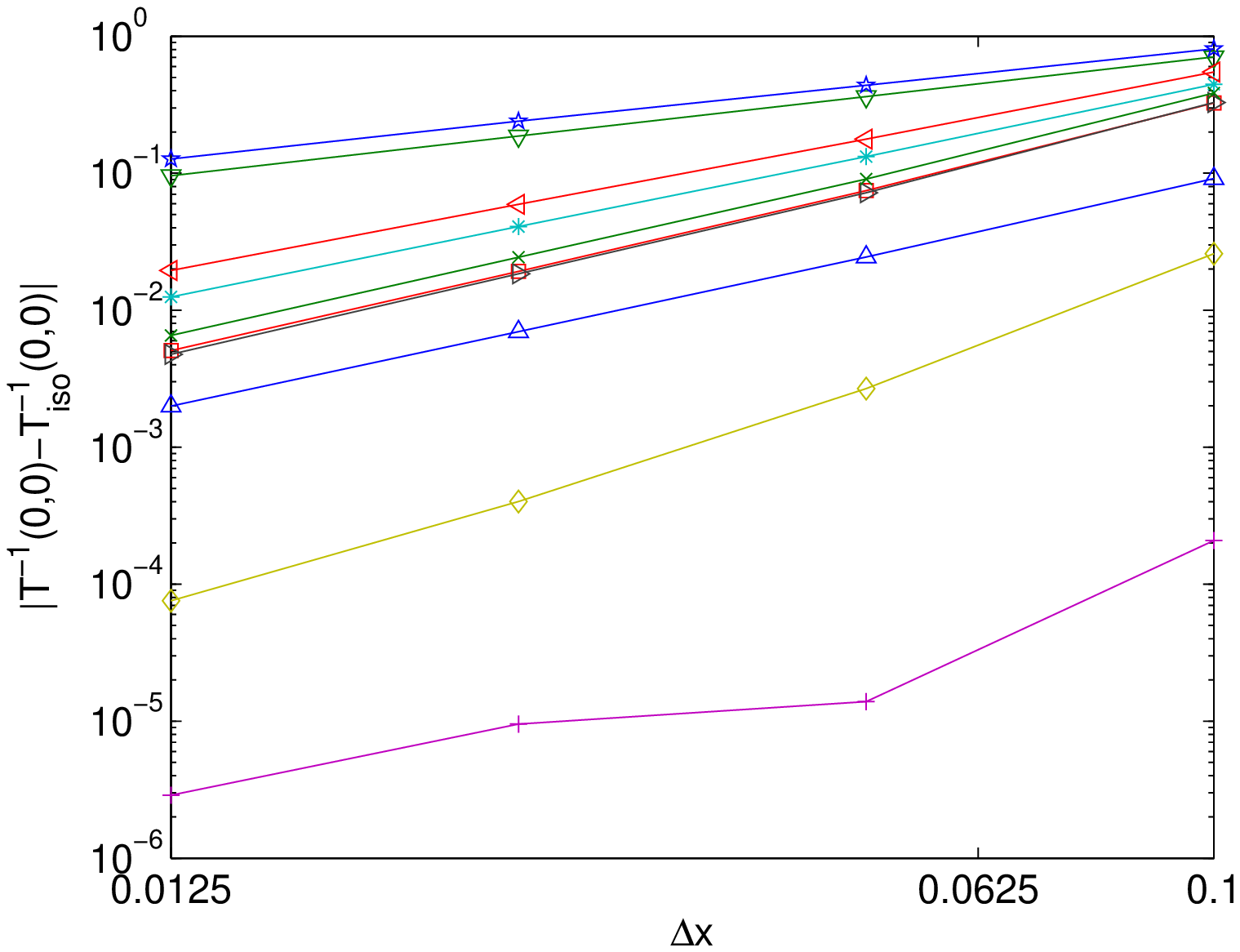}
\includegraphics[width=4 in, height=3 in]{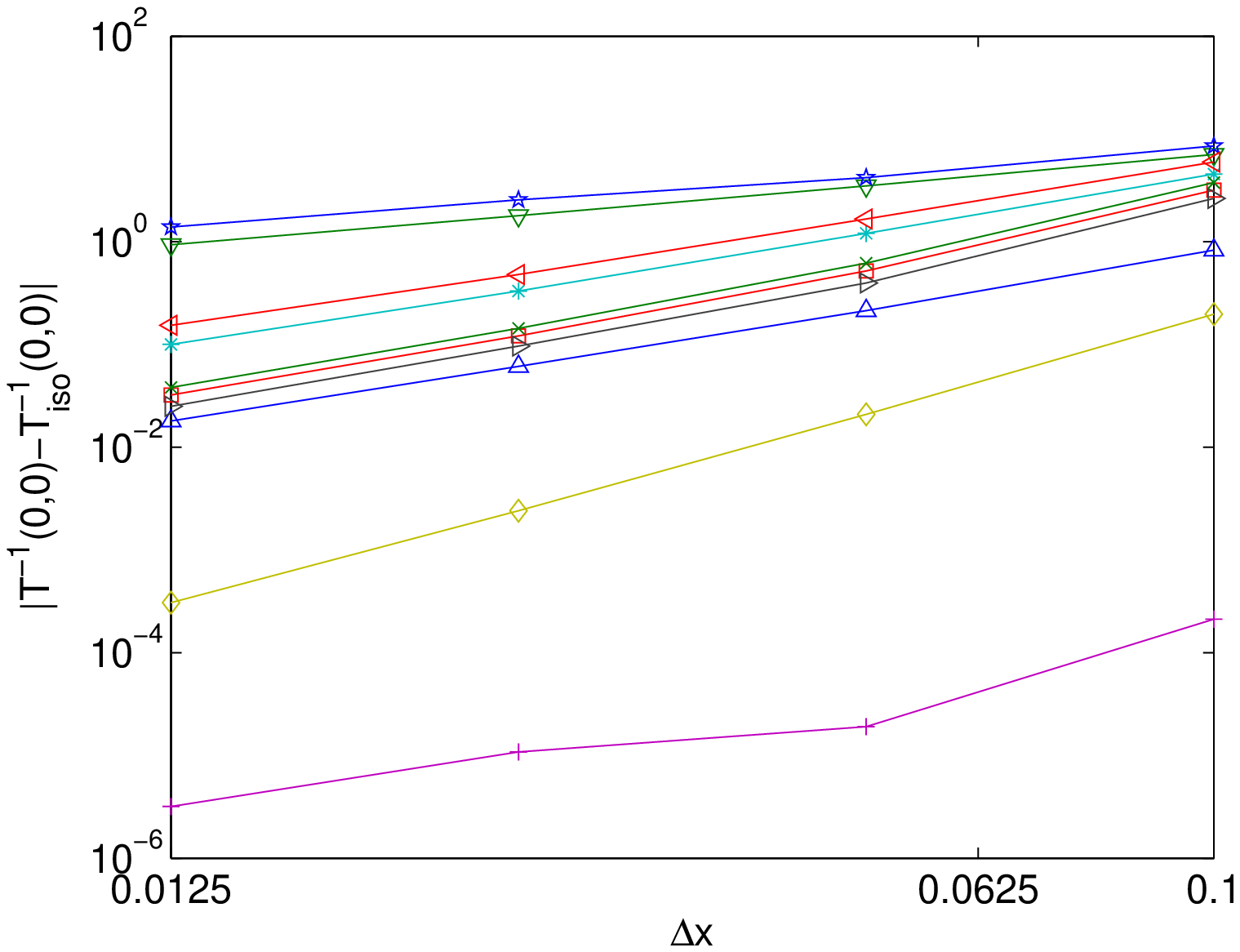}
\caption{A measure of perpendicular numerical diffusion
$\chi_{\perp,{\rm num}} = |T^{-1}(0,0)-T^{-1}_{\rm iso}|$ for
$\chi_\parallel/\chi_\perp=10$ (top) and
$\chi_\parallel/\chi_\perp=100$ (bottom), using different
methods. The different schemes are:
asymmetric ($\triangle$), asymmetric with minmod ($\triangledown$),
asymmetric with MC ($\square$), asymmetric
with van Leer ($\ast$), symmetric ($+$), symmetric with entropy
limiting ($\diamond$), symmetric with entropy and extrema limiting
($\triangleright$), symmetric with minmod ($\star$), symmetric with
MC ($\times$), and symmetric with van Leer limiter
($\triangleleft$). The numerical diffusion scales with
$\chi_\parallel$ for all methods except the symmetric differencing
\cite{Gunter2005}.
\label{fig:fig9}}
\end{figure}

We use the steady state test problem described in
\cite{Sovinec2004} to measure the perpendicular numerical diffusion
coefficient, $\chi_{\perp,{\rm num}}$. The computational domain is
a unit square $[-0.5,0.5]\times[-0.5,0.5]$, with vanishing temperature at the
boundaries; number density is set to unity.
The source term $Q=2\pi^2 \cos(\pi x) \cos(\pi y)$ that drives the
lowest eigenmode of the temperature distribution is added to Eq.
(\ref{eq:anisotropic_conduction}). The anisotropic diffusion
equation with a source term possesses a steady state solution. The
equation that we evolve is \be
\label{eq:anisotropic_conduction_source} \frac{\partial
e}{\partial t} = - \vec{\nabla} \cdot \vec{q} + Q .\ee

The magnetic field is derived from the flux function of the form
$\psi \propto \cos(\pi x)$ $\cos(\pi y)$; this results in concentric field
lines centered at the origin.
The temperature eigenmode driven by
the source function $Q$ is constant along the field lines. The
steady state solution for the temperature is $T(x,y)=\chi_\perp^{-1}
\cos(\pi x) \cos(\pi y)$, independent of $\chi_\parallel$. The
perpendicular diffusion coefficient $\chi_\perp$ is chosen to be
unity, thus $T^{-1}(0,0)$ gives a measure of total perpendicular
diffusion: the sum of $\chi_\perp$ (the explicit perpendicular
diffusion) and $\chi_{\perp, {\rm num}}$ (the perpendicular numerical
diffusion).

To account for $\chi_{\perp, {\rm num}}$ due to the errors in discretization
of the parallel diffusion operator, we calculate
$\chi_{\perp, {\rm num}} = |T^{-1}(0,0) - T^{-1}_{\rm iso}(0,0)|$, where
$T_{\rm iso}(0,0)$ is the central temperature calculated by the
discretized equations at the
same resolution in the isotropic limit $\chi_\parallel=\chi_\perp$.
The convention that we use is slightly different (and more accurate) than
that used in previous work, $\chi_{\perp, {\rm num}} = |T^{-1}(0,0) - 1|$,
which effectively assumed that
isotropic diffusion gives $T_{\rm iso}(0,0)=1$ exactly.

Figure \ref{fig:fig9} shows the perpendicular numerical diffusivity
$\chi_{\perp,{\rm num}}= |T^{-1}(0,0)-T^{-1}_{\rm iso}(0,0)|$ for
$\chi_\parallel/\chi_\perp=10$, $100$ using different methods. The
perpendicular diffusion ($\chi_{\perp,{\rm num}}$) for all methods
except the symmetric method increases linearly with
$\chi_\parallel$. This property has been emphasized by
\cite{Gunter2005} to motivate the use of symmetric differencing for
fusion applications, which require the perpendicular numerical
diffusion to be small for $\chi_\parallel/\chi_\perp \sim 10^9$. The
slope limited methods (with a reasonable resolution) are not
suitable for the applications which require
$\chi_\parallel/\chi_\perp \gg 10^4$; this rules out the fusion
applications mentioned in \cite{Gunter2005,Sovinec2004}. However,
only the slope limited methods give physically appropriate behavior
at temperature extrema, thereby avoiding negative temperatures in
presence of sharp temperature gradients. The error (perpendicular
numerical diffusion, $\chi_{\perp,{\rm
num}}=|T^{-1}(0,0)-T^{-1}_{\rm iso}(0,0)|$) for most methods except
the ones based on minmod limiter, show a roughly second order
convergence (see Table \ref{tab:tab6}).

\begin{table}[hbt]
\centering
\caption{Asymptotic slopes for convergence of error
$\chi_{\perp,{\rm num}} = |T^{-1}(0,0)-T^{-1}_{\rm iso}(0,0)|$ \label{tab:tab6}}
\begin{tabular}{ccc}
\hline
Method & $\chi_\parallel/\chi_\perp=10$ & $\chi_\parallel/\chi_\perp=100$ \\
\hline
asymmetric & 1.802  & 1.770 \\
asymmetric minmod & 0.9674 & 0.9406 \\
asymmetric MC & 1.9185 & 1.9076 \\
asymmetric van Leer & 1.706 & 1.728 \\
symmetric & 1.726 & 1.762 \\
symmetric entropy & 2.407 & 2.966 \\
symmetric entropy extrema & 1.949 & 1.953 \\
symmetric minmod & 0.9155 & 0.8761 \\
symmetric MC & 1.896 & 1.9049 \\
symmetric van Leer & 1.6041 & 1.6440 \\
\hline
\end{tabular}
\end{table}

\section{Conclusions}
It is shown that simple centered differencing of anisotropic
conduction can result in negative temperatures in the presence of large
temperature gradients. We present simple test problems where
asymmetric and symmetric methods give heat flowing from
lower to higher temperatures, leading to negative temperatures at
some grid points. Negative temperature results in numerical
instabilities, as the sound speed becomes imaginary. Numerical
schemes based on slope limiters are proposed to solve this problem.

The methods developed here will be useful in numerical studies of
hot, dilute, anisotropic astrophysical plasmas
\cite{Parrish2005,Sharma2006}, where large temperature gradients may
be common. Anisotropic conduction can play a crucial role in
determining the global structure of hot, non-radiative accretion
flows (e.g., \cite{Balbus2001,Parrish2005,Sharma2006}). Therefore,
it will be useful to extend ideal MHD codes used in previous global
numerical studies (e.g., \cite{Stone2001}) to include anisotropic
conduction. Slope limiting methods that prevent negative temperature
can be particularly helpful in global disk simulations where there
are huge temperature gradients that occur between a hot, dilute
corona and the cold, dense disk. The slope limited method with an MC
limiter appears to be the most accurate method that does not result
in unphysical behavior with large temperature gradients (see Figures
\ref{fig:fig6} \& \ref{fig:fig8}).
While we have tried a number of possible variations other than the ones
described here, there might be ways to further improve these algorithms.
Future work might explore other combinations of limiters, or limiters on
combined fluxes instead of limiting the normal and transverse components
independently, or might explore using higher-order information to reduce
the effects of limiters near extrema while preserving physical behavior.

Although the slope and entropy limited methods in the present form
are not suitable for fusion applications that require accurate
resolution of perpendicular diffusion for huge anisotropy
($\chi_\parallel/\chi_\perp \sim 10^9$), they are appropriate for
astrophysical applications with large temperature gradients. A
relatively small anisotropy of thermal conduction may be sufficient to
study the effects of anisotropic thermal conduction~\cite{Parrish2005}.
The primary advantage of the limited methods is their robustness in
presence of large temperature gradients. Apart from the simulations
of dilute astrophysical plasmas with large temperature gradients
(e.g., magnetized collisionless shocks), monotonicity-preserving methods may find
use in diverse fields where anisotropic diffusion is important,
e.g., image processing, biological transport, and geological systems.

\section{Acknowledgments}
Useful discussions with Tom Gardiner, Ian Parrish, and
Ravi Samtaney are acknowledged. This work is supported by the US DOE under
contract \# DE-AC02-76CH03073 and the NASA grant NNH06AD01I.

\appendix \section{Entropy condition for an ideal gas}
\label{app:app5} The entropy for an ideal gas is given by $S =
nVk \ln(T^{1/(\gamma-1)}/n) + \mbox{const.}$, where $n$ is the number
density, V the volume, $T$ the temperature, and $\gamma$ the ratio
of specific heats ($=5/3$ for a 3-D mono-atomic gas). The change in
entropy that results from adding an amount of heat $dQ$ to a uniform gas is
$$
d S = \frac{n V k}{\gamma - 1} \frac{dT}{T} = \frac{dQ}{T}.
$$
We measure temperature in energy units, so $k=1$.
The
rate of change of entropy of a system where number density and
temperature can vary in space (density is assumed to be constant in
time) is given by \be \dot{S} \equiv \frac{\partial S}{\partial t} =
- \int dV \frac{ \vec{\grad} \cdot \vec{q}}{T} = - \int dV
\frac{\vec{q} \cdot \vec{\grad} T}{T^2} = \int dV n \chi
\frac{|\nabla_\parallel T|^2}{T^2} \geq 0, \ee where we use an
anisotropic heat flux, $\vec{q}=-n\chi \vec{b}\vec{b} \cdot
\vec{\grad} T $, and the integral is evaluated over the whole space
with the boundary contributions assumed to vanish. The local entropy
function defined as $\dot{s}=-\vec{q} \cdot \vec{\grad}T/T^2$ can be
integrated to calculate the rate of change of total entropy of the
system.

In the paper we use a related function (the entropy-like function
$\dot{s}^*$) defined as $\dot{s}^* \equiv -\vec{q} \cdot
\vec{\grad}T$ to limit the symmetric methods using face-pairs, and
to prove some properties of different anisotropic diffusion schemes.
The condition $-\vec{q} \cdot \vec{\grad}T \geq 0$ ensures that heat
always flows from higher to lower temperatures.

\end{document}